\documentclass[journal,10pt]{IEEEtran}
\usepackage[utf8]{inputenc}
\usepackage{amsmath}
\usepackage{amssymb}
\usepackage{amsthm,color}
\usepackage{amsmath}
\usepackage{caption}
\usepackage{multirow}
\usepackage{subcaption}
\usepackage{graphicx} 
\usepackage{epstopdf}
\usepackage[rightcaption]{sidecap}

\usepackage{tikz}
\usetikzlibrary{positioning, shapes.geometric, arrows.meta}
\usetikzlibrary{arrows,decorations.pathreplacing}

\newtheorem{theorem}{Theorem}
\newtheorem{property}{Property}
\newtheorem{corollary}{Corollary}
\newtheorem{definition}{Definition}

\newtheorem{lemma}{Lemma}
\newcommand{\oneton}{1,\cdots,n}
\newcommand{\R}{\mathbb R}

\newcommand{\V}{\mathcal V}
\newcommand{\W}{\mathcal W}
\newcommand{\N}{\mathbb N}

\newcommand{\G}{\mathcal G}
\newcommand{\F}{\mathcal F}
\newcommand{\bo}{\backslash\{\mathbf{0}\}}
\graphicspath{{figs/}}
\usepackage[square,numbers]{natbib}
\newcommand{\ignore}[1]{}

\begin{document}
\title{Preventive and Reactive Cyber Defense Dynamics with Ergodic Time-dependent Parameters Is Globally Attractive}
\author{Yujuan Han, Wenlian Lu, Shouhuai Xu
\thanks{Y. Han is with College of Information Engineering, Shanghai Maritime University, China and Department of Computer Science, University of Texas at San Antonio, USA. W. Lu is with School of Mathematical Sciences, Fudan University, China; Shanghai Center for Mathematical Sciences, Fudan University, China; and Shanghai Key Laboratory for Contemporary Applied Mathematics, China. S. Xu is with Department of Computer Science, University of Texas at San Antonio, USA. Correspondence: shxu@cs.utsa.edu.}
}

\IEEEtitleabstractindextext{
\begin{abstract}
Cybersecurity dynamics is a mathematical approach to modeling and analyzing cyber attack-defense interactions in networks. In this paper, we advance the state-of-the-art in characterizing one kind of cybersecurity dynamics, known as {\em preventive and reactive cyber defense dynamics}, which is a family of highly nonlinear system models. We prove that this dynamics in its general form with {\em time-dependent} parameters is {\em globally attractive} when the time-dependent parameters are {\em ergodic}, and is {\em (almost) periodic} when the time-dependent parameters have the stronger properties of being {\em (almost) periodic}. Our results supersede the state-of-the-art ones, including that the same type of dynamics but with {\em time-independent} parameters is {\em globally convergent}. 
\end{abstract}

\begin{IEEEkeywords}
Cybersecurity dynamics, preventive and reactive cyber defense dynamics, global attractivity, network science
\end{IEEEkeywords}}

\maketitle

\IEEEpeerreviewmaketitle

\IEEEdisplaynontitleabstractindextext

\section{Introduction}\label{sec:introduction}

Cyberspace is a complex system that has become a critical infrastructure. However, our understanding of its security is still superficial, explaining why there are so many cyber attacks on a daily basis. This calls for research in understanding cybersecurity at many levels of abstractions, ranging from macroscopic to microscopic \cite{XuCybersecurityDynamicsHotSoS2014,XuEmergentBehaviorHotSoS2014,XuCDBookChapter2019}. At a macroscopic level, Kephart and White \cite{KephartOkland91,KephartOkland93} adapt the classic biological epidemic models \cite{Kermack1927} to cyberspace, while inheriting the homogeneity assumption that each node in a network has equal chance in attacking any other node. This approach is later extended to accommodate heterogeneous network structures represented by {\em arbitrary} adjacency matrices \cite{WangSRDS03}. 

These studies \cite{KephartOkland91,KephartOkland93,WangSRDS03} have inspired a systematic approach, dubbed Cybersecurity Dynamics 
\cite{XuCybersecurityDynamicsHotSoS2014,XuEmergentBehaviorHotSoS2014,XuCDBookChapter2019}, which can be characterized as follows.
First, it proposes using arbitrary matrices to describe the {\em attack-defense} structures that are induced by cybersecurity (e.g., access control) policies enforced on top of physical networks. That is, adjacency matrices are used to model which nodes are allowed to communicate with which other nodes through some routing paths (with each typically consisting of multiple point-to-point physical communication links), rather than modeling the physical links. 

Second, computer epidemic models, including \cite{KephartOkland91,KephartOkland93,WangSRDS03,TowsleyInfocom05} and their numerous follow-up studies, often focus on investigating {\em epidemic threshold}, which distinguishes the parameter regime in which the spreading dies out from the parameter regime in which the spreading doesn't. While important, this understanding is far from sufficient  in cybersecurity. For example, we need to know whether the spreading will be converging or not when it does not die out. 

Third, the rich semantics of cyber attack-defense interactions has led to families of cybersecurity dynamics models, including: preventive and reactive cyber defense dynamics \cite{XuAINA07,XuTDSC2011,XuTAAS2012,zheng2018preventive,lin2018unified}, active cyber defense dynamics \cite{XuGameSec13,XuInternetMath2015ACD,XuHotSoS15}, adaptive cyber defense dynamics \cite{XuQuantitativeSecurityHotSoS2014,XuTAAS2014}, and proactive cyber defense dynamics \cite{XuHotSOS14-MTD}. These theoretical studies have deepened our understanding of cybersecurity. For example, now we know: preventive and reactive cyber defense dynamics is {\em globally convergent} in certain settings \cite{zheng2018preventive,lin2018unified}, and {\em global convergence} is a nice cybersecurity property that makes it possible to predict and manage cybersecurity \cite{XuTAAS2012}; in contrast, active cyber defense dynamics can be Chaotic \cite{XuHotSoS15}. 

In this paper, we focus on the aforementioned preventive and reactive cyber defense dynamics, which is a family of highly nonlinear system models that are initiated in \cite{XuAINA07} and inspired by \cite{WangSRDS03}. This dynamics aims to model the interactions between two classes of cyber attacks and two classes of cyber defenses. The two classes of attacks are: {\em push-based} attacks (including computer malware spreading) and {\em pull-based} attacks (including “drive-by download” attacks; i.e., a computer gets compromised when visiting a malicious website \cite{ProvosHotbot07}). The two classes of defenses are: {\em preventive} defenses, which include the use of access control and intrusion prevention mechanisms to attempt to prevent attacks from succeeding; and {\em reactive} defenses, which include the use of anti-malware and intrusion detection mechanisms to attempt to detect and clean up compromised computers. Since this dynamics uses a certain {\em product} term, which will be elaborated later, to model the collective effect on a node when attacked by others, it is also known as the $\prod$-model.

Closely related to the $\prod$-model is the the $N$-intertwined model \cite{VanMieghemIEEEACMTON09}, or the $\sum$-model because it uses a certain {\em additive} term (which will be elaborated later as well) to model the collective effect on a node when attacked by others. This model is also inspired by \cite{WangSRDS03} and has been studied in, for example,  \cite{VanMieghem:2011:NIS:2141982.2141984,Chai:2017:PES:3068707.3068740,AAFALL2007-their-80,6859418--their-76,Sahneh:2013:GEM:2578911.2578931}.

A more general preventive and reactive cyber defense dynamics is introduced in  \cite{lin2018unified}, which accommodates the aforementioned $\prod$-model and $\sum$-model as two special cases and is thus dubbed the {\em unified dynamics} or {\em unified model}. A remarkable result is that the unified model 
is globally convergent in the {\em entire} parameter universe \cite{lin2018unified}. As shown in \cite{lin2018unified}, this result supersedes many results presented in the literature, which often deal with some special cases of the unified model. 
While fairly general, this unified model  \cite{lin2018unified} only accommodates {\em time-independent} parameters (i.e., parameter values do not change over time). 
This {\em time-independence} of parameters is restrictive, and should be eliminated to accommodate {\em time-dependent} parameters to make the theoretical results more widely applicable. This motivates the present study.

\subsection{Our Contributions}

\begin{figure}[!htbp]
	\centering
	\begin{subfigure}[b]{0.2\textwidth}
		\centering
		\includegraphics[width=\textwidth]{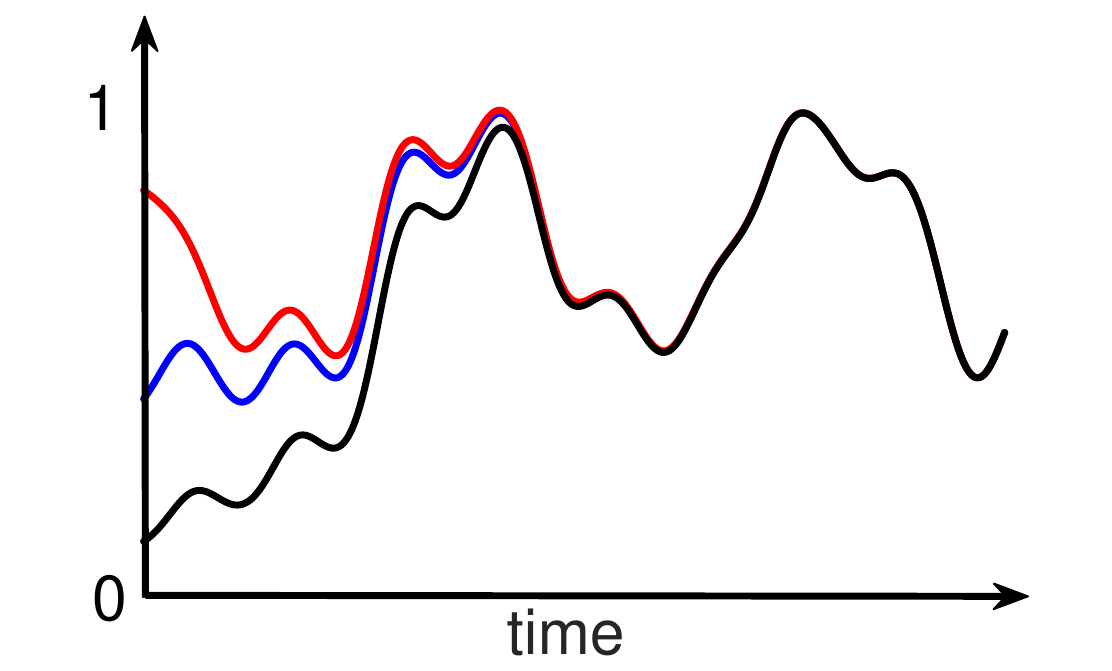}
		\caption{{\footnotesize Global attractivity}}
		\label{fig:global-attractive}
	\end{subfigure}
	\begin{subfigure}[b]{0.2\textwidth}
		\centering
		\includegraphics[width=\textwidth]{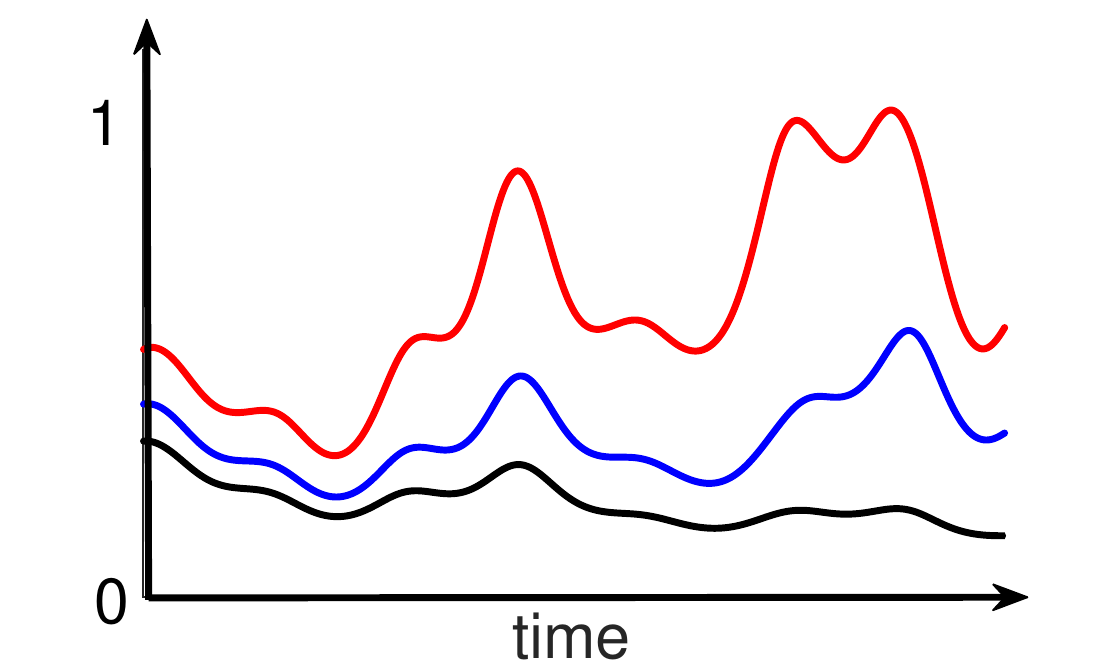}
		\caption{{\footnotesize Non-global attractivity}}
		\label{fig:global-unstable}
	\end{subfigure}
\caption{Illustration of  global attractivity vs. non-global attractivity, where the latter is sensitive to the initial value and may be even Chaotic and therefore unpredictable.}
\label{fig:global-attractive-illstrate}
\end{figure}

We investigate preventive and reactive cyber defense dynamics with 
{\em time-dependent} parameters, or the {\em unified model with time-dependent parameters}. 
Our results supersede the ones obtained in the unified model with time-independent parameters  \cite{lin2018unified};
our results also supersede previous results that are obtained in special cases of the time-dependent $\prod$-model \cite{XuTAAS2014,zhang2017spectral,prakash2010virus,sanatkar2016epidemic} and the time-dependent $\sum$-model \cite{bokharaie2010spread,pare2018epidemic}.
All of these results typically deal with convergence to an equilibrium, meaning that their techniques are no longer applicable in our setting,
explaining why we adopt the skew-product semi-flow approach and the multiplicative ergodic theorem in the present paper. We prove that the unified model 
with {\em ergodic} parameters is globally attractive (i.e., converging to a unique trajectory regardless of the initial value
as illustrated in Figure \ref{fig:global-attractive-illstrate}(a) where the $y$-axis represents a metric of interest (e.g., the fraction of compromised nodes in a network \cite{XuACMComputingSurvey2017}), whereas Figure \ref{fig:global-attractive-illstrate}(b) illustrates the absence of global attractivity.
We further prove that when the parameters are almost periodic (vs. periodic), the unified model with time-dependent parameters is also almost periodic (correspondingly, periodic). In addition, we present bounds on the globally attractive trajectory, which are useful because (for example) the upper bound can be seen as the worst case scenario for decision-making purposes when only partial information about the parameters is given.

\begin{SCfigure*}[1][!htbp]
\caption{Illustrating how our results (concerning the unified model with {\em time-dependent} parameters) supersede:  (i) the results of the unified model with {\em time-independent} parameters presented in \cite{lin2018unified}, which supersede numerous results respectively obtained in the $\prod$-model and $\sum$-model (cf. \cite{lin2018unified} for details); and (ii) the results on the convergence to equilibrium $\mathbf{0}$ of the $\prod$-model and $\sum$-model with {\em time-dependent} parameters. Note that (i) and (ii) intersect in the special case the $\prod$-model and $\sum$-model with {\em time-independent} parameters converge to  $\mathbf{0}$.}
{
\scalebox{0.8}
{
\begin{tikzpicture}
\draw[very thick] (-5.8,1.4) rectangle (6.3,7.1);
\draw[very thick] (-2.25,5) ellipse (3.2cm and 1.9cm);
\draw[very thick] (2.7,5) ellipse (3.2cm and 1.9cm);
\node(tx1) at (-2.25,5)  [text width=3.9cm]	{\textbf{Known:} (i) In {\em time-independent} parameters setting, the {\em unified model} (accommodating the $\prod$-model and the $\sum$-model as special cases) 
is globally convergent \cite{lin2018unified}.};
\node(tx2) at (3.1,4.98)  [text width=4cm]	{\textbf{Known:} (ii) In {\em time-dependent} parameters setting, the $\prod$-model \cite{XuTAAS2014,zhang2017spectral,prakash2010virus,sanatkar2016epidemic} 
and $\sum$-model  \cite{bokharaie2010spread,pare2018epidemic} 
globally converge to $\mathbf{0}$ under certain conditions when there are no pull-based attacks.};
\node(tx3) at (0.25,2.2)  [text width=11.7cm]	{{\bf Our results}: The unified model with {\em time-dependent} parameters  is globally attractive when the parameters are \textit {ergodic} (Theorem \ref{thm-general}),
and globally converges to $\mathbf{0}$ under a certain condition 
when there are no pull-based attacks
(Theorem \ref{thm-alpha-0-stability}).};
\end{tikzpicture}
}
}
\label{fig:illustration}
\end{SCfigure*}

In order to systematize knowledge, we use Figure \ref{fig:illustration} to highlight how the present paper supersedes the literature results.
Our Theorem \ref{thm-general} supersedes the global convergence of the  unified model with {\em time-independent} parameters presented in \cite{lin2018unified}, which in turn supersedes numerous prior results reported in the literature; this is because the ergodicity condition (or assumption) required by Theorem \ref{thm-general} naturally holds in the time-independent setting. When there are no pull-based attacks (i.e., considering push-based attacks only), our Theorem \ref{thm-alpha-0-stability} shows that under even weaker condition (than ergodicity) the unified model with {\em time-dependent} parameters globally converge to the special equilibrium $\mathbf{0}$ (i.e., there are no compromised nodes); this supersedes the literature results 
on the $\prod$-dynamics and $\sum$-dynamics with {\em time-dependent} parameters, because the condition required by our Theorem \ref{thm-alpha-0-stability} naturally holds in the $\prod$-model and $\sum$-model with time-dependent parameters investigated in the literature.

\subsection{Related Work}
\label{sec-related-works}

To the best of our knowledge, there is no prior study that aims to characterize the unified model with time-dependent parameters in the entire parameter universe. Prior studies related to preventive and reactive cyber defense dynamics can be divided into two categories:
considering {\em time-independent} parameters vs. considering {\em time-dependent} parameters. For {\em time-independent} models,
the state-of-the-art is the global convergence result of the unified model \cite{lin2018unified}, which supersedes numerous results in the $\prod$-model (e.g., \cite{XuTAAS2012,zheng2018preventive,lin2018unified}) and the $\sum$-model (e.g., \cite{VanMieghemIEEEACMTON09,VanMieghem:2011:NIS:2141982.2141984,Chai:2017:PES:3068707.3068740,AAFALL2007-their-80,6859418--their-76,Sahneh:2013:GEM:2578911.2578931}). 
Time-dependent parameters have been investigated in the $\prod$-model and $\sum$-model, but only to the extent of understanding when the dynamics converges to the special equilibrium $\mathbf{0}$ while assuming there are no pull-based attacks. Specifically, \cite{XuTAAS2014} investigates the $\prod$-model with time-dependent parameters and identifies a sufficient condition under which the dynamics converges to  $\mathbf{0}$; 
\cite{zhang2017spectral,prakash2010virus,sanatkar2016epidemic} investigate the convergence to  $\mathbf{0}$ of the $\prod$-model with time-independent parameters. For the $\sum$-model with time-dependent parameters while assuming there are no pull-based attacks, sufficient conditions under which the dynamics converges to  $\mathbf{0}$ are presented in \cite{bokharaie2010spread,pare2018epidemic,rami2014stability,ogura2015disease}. 

\subsection{Paper Outline}
Section \ref{sec-pre} reviews some mathematical knowledge. Section \ref{sec-new-results} presents and investigates the unified model with time-dependent parameters. Section \ref{sec-numerical} reports our simulation study. Section \ref{sec-conclusion} concludes the paper with open problems.

\section{Mathematical Preliminaries}\label{sec-pre}

Let $\R$ be the set of real numbers, $\R^n_{\ge 0} = \{x=(x_1,\ldots,x_n) \in \R^n: 
x_v\ge 0 ~{\rm for}~ 1\leq v\leq n\}$, $\R^n_{> 0} = \{x\in\R^n_{\ge 0}: x_v> 0 ~{\rm for ~some}~ 1\leq v\leq n\}$, $\R^n_{\gg 0} = \{x\in\R^n_{>0}: \forall v,~x_v> 0\}$, $[0,1]^n = \{x\in\R^n: 0 \le x_v\le 1\}$, and $[0,1]^{n,n} = \{A\in\R^{n\times n}: 1 \le A_{uv}\le 0\}$. Let $I_n$ denote the $n\times n$ identity matrix. For a matrix $A=(a_{vu})$, let $A\ge \delta$ denote that every element of $A$ is greater than $\delta$, namely $a_{vu}\ge \delta$~ $\forall v,u$.
For two vectors $x,z\in\R^n$, let $x\ge z$ denote $x-z\in \R^n_{\ge 0}$, $x>z$ denote $x-z\in \R^n_{> 0}$, and $x\gg z$ denote $x-z\in \R^n_{\gg 0}$. 
Table \ref{table:model-variables-and-parameters} summarizes the other notations.

\begin{table}[!t]
	\centering 
	\caption{Notations used in the paper.\label{table:model-variables-and-parameters}}
	\begin{tabular}{|l|p{6.4cm}|}
		\hline
$G(t)$, $A(t)$ & attack-defense structure $G(t)=(V,E(t))$ where $V=\{1,\cdots,n\}$ is the set of nodes and $E(t)$ is the set of arcs; its adjacency matrix $A(t)=[a_{vu}(t)]_{n\times n}$ where $a_{vu}(t)=1$ if and only if $(u,v)\in E(t)$ \\  \hline
$N_{v}(t)$ & $v$'s neighbors that are allowed to communicate with $v$ at time $t$, i.e., $N_{v}(t)=\{u\in V:(u,v)\in E(t)\}$\\ \hline
$\gamma_{vu}(t)$, $\Gamma(t)$ & $\gamma_{vu}(t)\in [0,1]$ is the probability a {\em secure} node $v\in V$ becomes {\em compromised} because of the push-based attack waged by {\em compromised} neighbor $u\in N_v(t)$; $\Gamma(t)=[\gamma_{vu}(t)]_{n\times n}$ \\ \hline 
$\mu(C(t))$ &  the maximum Lyapunov exponent (MLE) of system $d{z}(t)/dt= C(t)z(t)$; $\mu(C(t))= \limsup_{t\to\infty}\frac{1}{t}\log\|U(t,0)\|$, where $U(t,0)$ is the {\em fundamental solution matrix} of the system\\ \hline
$\lambda_1(B), \rho(B)$ &  the maximum eigenvalue of $B$ (in real part); the spectral radius of $B$\\ \hline
$[x]_{\V}$,$ [B]_{\V,\W}$  & for any vector $x\in\R^n$ and any index subset $\V \subseteq \{\oneton\}$, $[x]_{\V}=(x_v)_{v\in \V}\in\R^{|\V|}$;  for any matrix $B = [b_{vu}]\in\R^{n\times n}$ and two index subsets $\V, \W\subseteq \{\oneton\}$, $[B]_{\V,\W}=[b_{vu}]_{v\in\V,u\in\W}\in \R^{|\V|\times |\W|}$ \\ \hline
$\mathbf{0},\mathbf{0}_d$ &  $\mathbf{0}=[0,\dots,0]\in\R^n$, $\mathbf{0}_{d} = [0,\dots,0]\in\R^d$ \\ \hline
$i_v(t)$, $s_v(t)$ & the probability that node $v\in V$ is {\em compromised} and {\em secure} at time $t$, where $i_v(t)+s_v(t)=1$ \\ \hline
$\alpha_{v}(t)$, $\alpha(t)$ & $\alpha_{v}(t)\in [0,1]$ is the probability that a {\em secure} node $v\in V$ becomes {\em compromised} at time $t$ because of pull-based attacks; $\alpha(t)=[\alpha_1(t),\cdots,\alpha_n(t)]$\\ \hline
$\beta_{v}(t)$, $\beta(t)$ & $\beta_{v}(t)\in [0,1]$ is the probability that a {\em compromised} node $v\in V$ becomes {\em secure} at time $t$ because of the reactive defense; $\beta(t)=[\beta_1(t),\cdots,\beta_n(t)]$\\  \hline
$y(t)$  & $y(t) = [\alpha(t),\beta(t),{\rm vec}(A(t))^{\top},{\rm vec}(\Gamma(t))^{\top}]$ is the vector of model parameters (including the attack-defense structure), where ${\rm vec}(A(t))$ is the column vector obtained by stacking the columns of matrix $A(t)$\\ \hline
$D_x f(x,y)$ & $D_x f(x,y) = \left[\partial f_v(x,y)/\partial x_u\right]_{u,v\in V}$, the Jacobian matrix of $f:X\times Y\to X$ at point $(x,y)$ w.r.t. $x$ \\ \hline
$\mathcal M(\R,\R^m)$ & the space of measurable functions from $\R$ to $\R^m$\\ \hline
\end{tabular}
\end{table}

\subsection{Ergodicity and Almost Periodicity}
We propose modeling {\em time-dependent} parameters $\{y(t)\}_{t\ge 0}$ as a sequence drawn from an {\em ergodic} stochastic process because we can only observe a single sequence of a stochastic process in the real world. Therefore, it is necessary to require that the observed sequence is representative of the stochastic process, meaning that the averaged behavior of the sequence is the same as the average over the probability space. Formally,

\begin{definition}[ergodicity and mean value \cite{arnold2013random,birkhoff1931proof}]\label{def-ergodic}
	Let $Y\subset \mathcal M(\R,\R^m)$ be a compact space of measurable functions from $\R$ to $\R^m$, $(Y,\F,\mathbb P)$ be a probability space, and $\theta: \R^+\times Y\to Y$  be the right-shift translation that $\theta(s,y(t)) = y(t+s)$, $\forall s,t\ge 0$. Measure $\mathbb P$ is called
	{\em $\theta$-invariant} if for any $K\in \mathcal F$, $\mathbb P(K) = \mathbb P(\theta^{-1}(t)K)$ holds for all $t\in\R$,
	and called {\em $\theta$-ergodic} if for any $A\in \mathcal F$ that satisfies $A = \theta^{-1}(t)A$, $\forall t\in \R$, $A$ has measure 0 or 1. 
	We call $y \in Y$, namely $\{y(t)\}_{t\ge 0} \in Y$, {\em ergodic} if and only if $\mathbb P$ is {\em $\theta$-invariant} and {\em $\theta$-ergodic}. 
	
	For any ergodic process $\{y(t)\}_{t\ge 0}$, $M(y)$ is called the {\em mean value} of $\{y(t)\}_{t\ge 0}$, where  
	\begin{align}\label{eq:ergodic-mean}
	M(y) = \lim_{T\to\infty}\frac{1}{T}\int_{a}^{a+T}y(\tau)d\tau
	\end{align}
	holds uniformly with respect to any $a$ almost surely.
\end{definition}

Many stochastic processes are ergodic, such as a sequence of independent and identically distributed random variables and ergodic Markov processes \cite{borovkov1998ergodicity}. Moreover, {\em (almost) periodic} functions are also ergodic \cite{arnold2013random}.
Formally, we have:
\begin{definition}[almost periodicity \cite{besicovitoh1932almost}]
	\label{def-almost-periodic}
	A continuous function $y(t):\R \to \R^m$ is said {\em almost periodic} in $t$ if for any $\epsilon>0$, there exists a number $l(\epsilon)>0$ such that every interval of length $l(\epsilon)$ contains a point  $\xi\in \R$ and  
	\begin{align*}
	\|y(t+\xi)-y(t)\|<\epsilon,~\forall t\in\R,
	\end{align*}
	where $\xi$ is called ``{an $\epsilon$-translation number of $y(t)$}''. 
	A continuous function $f(t,x):\R\times X\to \R^n$ is called {\em almost periodic} in $t$ if for any $x\in X$, $f(t,x)$ is {\em almost periodic} in $t$. 
\end{definition}

An almost periodic time-dependent parameter $g(t)$, if not periodic, has no period, meaning $g(t+\tau)=g(t), \forall t\in\R$ does {\em not} hold for any $\tau\in\R$, but $g(t+\tau)\approx g(t), \forall t\in\R$ holds with any degree of approximation for infinitely many $\tau\in\R$, where $\tau$ may be very large. For example, $g(t) = 0.2\times \sin(\pi t) + 0.1\times \sin(2\sqrt{2} \pi t)+ 0.3$, which contains three periodic terms, is almost periodic but not periodic.

\subsection{Subhomogeneous and Cooperative Dynamical Systems}

We will take advantage of {\em cooperative} and {\em subhomogeneous} dynamical systems, where the former means that all of the off-diagonal terms of the Jacobian matrix of a dynamical system are non-negative and the latter (or sublinearity) is a generalization of concavity.

\begin{definition}
	[cooperative dynamical system \cite{HirschPaper1985}]
	\label{def-cooperative}
	Consider a region $X\subseteq \R^n_{\ge 0}$, a subspace $Y\subseteq \mathcal M\left(\R,\R^m_{\ge 0}\right)$,  $x=[x_1,\cdots,x_n]^{\top}\in X$, $y \in Y$, and $f(x,y)=[f_1(x,y),\cdots,f_n(x,y)]^{\top}: X\times Y \to X$. A nonautonomous system \begin{align}\label{coop-sys}
	\frac{dx}{dt} = f(x,y(t))
	\end{align}
	is said to be {\em cooperative} if $\partial f_v(x,y)/\partial x_u \ge 0$ holds for $\forall u \neq v$ and $\forall (x,y)\in X\times Y$.
\end{definition}

\begin{definition}[subhomogeneity and monotonicity \cite{ZhaoBook2003}] \label{def-subhomogeneous}
	A continuous map $f(x,y): X\times Y\to X$ is said to be 
	\begin{itemize}
		\item {\em subhomogeneous} if $f(\eta x,y)\ge \eta  f(x,y)$ holds for any $x\in X$, $y\in Y$, and $\eta \in(0,1)$.
		\item {\em strictly subhomogeneous} if $f(\eta  x,y)>\eta f(x,y)$ holds for any $x\in X$ with $x\gg 0$, any $y\in Y$ and $\eta \in(0,1)$.
		\item {\em strongly subhomogeneous} if $f(\eta  x,y)\gg\eta f(x,y)$ holds for any $x\in X$ with $x\gg 0$, any $y\in Y$ and $\eta \in(0,1)$.
		\item {\em monotone} if $f(x_1,y)\ge f(x_0,y)$ holds for any $x_1\ge x_0$ and $y\in Y$.
		\item {\em strictly monotone} if $f(x_1,y)>  f(x_0,y)$ holds for any $x_1> x_0$ and $y\in Y$.
		\item {\em strongly monotone} if $f(x_1,y)\gg f(x_0,y)$ holds for any $x_1> x_0$ and $y\in Y$.
	\end{itemize}
\end{definition}

\subsection{Globally Attractive Dynamical Systems}

The concept of {\em uniform persistence} describes the behavior that trajectories are eventually uniformly away from the boundary of some closed invariant subset \cite{freedman1994uniform, butler1986uniformly}. Intuitively, when the origin $\mathbf{0}$ is on the boundary,  uniform persistence implies the instability of the origin $\mathbf{0}$.

\begin{definition}[persistence and uniform persistence \cite{butler1986uniformly}]\label{def-persistence}
	Consider a closed region $X\subset \R^n_{\ge 0}$ with $\mathbf{0}\in X$ and a function space $Y\subset \mathcal M(\R,\R^m)$. For given $x_0\in X\bo$ and $y_0\in Y$, a continuous map $\psi(t,x_0,y_0): \R\to X $ is said to be 
	\begin{itemize}
		\item {\em persistent} if $\lim\inf_{t\to+\infty} \psi(t,x_0,y_0)\gg\mathbf{0}$ holds.
		\item {\em uniformly persistent} if $\lim\inf_{t\to+\infty} \psi(t,x_0,y)\gg\mathbf{0}$ holds for $\forall x_0\in X\bo$ and $\forall y\in Y$.
	\end{itemize}
\end{definition}

Denote by $\psi(t,x_0,y)$ the solution to system (\ref{coop-sys}) with respect to initial value $x(0) = x_0$ and $y\in Y$ (representing time-dependent parameters in the context of the present paper). Now  we introduce the the definition of {\em global attractivity}.

\begin{definition}[Global attractivity]
	\label{def-global-attractive}
	Consider system \eqref{coop-sys} with compact spaces $X\subseteq \R_{\ge 0}^n$ and $Y\subseteq \mathcal M(\R,\R^n)$, where $(Y,\mathcal F,\mathbb P)$ is a probability space and $y = \{y(t)\}_{t\ge 0}\in Y$ is a sample (or realization) of some time-dependent parameters. 
	\begin{itemize}
		\item[(i)] For system \eqref{coop-sys}, equilibrium $\mathbf{0}$ is said to be {\em almost surely globally attractive } if  $\lim_{t\to\infty} \psi(t,x_0,y)=\mathbf{0}$ holds for any $x_0\in  X$ and almost every $y\in Y$.
		\item[(ii)] For system \eqref{coop-sys} with a given $y\in Y$, a trajectory $\psi(t,x^*_y,y)$ is said to be {\em globally attractive} if $\lim_{t\to\infty} \|\psi(t,x_0,y)-\psi(t,x^*_y,y)\|= 0$ holds for any $x_0\in X\bo$.
		\item[(iii)] System \eqref{coop-sys} is said to be {\em almost surely globally attractive } if for almost every $y\in Y$, there exists a globally attractive trajectory $\psi(t,x^*_y,y)$, namely that $\lim_{t\to\infty} \|\psi(t,x_0,y)-\psi(t,x^*_y,y)\|= 0$ holds for any $x_0\in X\bo$.
	\end{itemize}
\end{definition}

Theorem \ref{thm-pre-global-attractivity} below bridges the preceding two concepts.

\begin{theorem}[Theorem 2.3.5 in \cite{ZhaoBook2003}] \label{thm-pre-global-attractivity}
	Let $X\subseteq \R_{\ge 0}^n$ and $Y$ be  compact spaces, where $Y$ has no nonempty, proper, closed invariant subset with respect to $\theta$ and there is a metric $d$ such that for any distinct $y_1, y_2\in Y$, $\inf_{t\in\R}d(\theta(t,y_1),\theta(t,y_2))>0$. Let $H: \R\times X\times Y\to X\times Y$ be the skew-product semi-flow associated to system \eqref{coop-sys} in the form 
	\begin{align*}
	H(t,x,y) = (\psi(t,x,y),\theta(t,y)),
	\end{align*} 
	where $\theta:\R\times Y\to Y$ is the right-shift translation (see Definition \ref{def-ergodic}). Denote by ${\rm int}(X) = X\cap \R_{\gg 0}^n$. Suppose 
	\begin{itemize}
		\item[(I)] for any $\eta\in(0,1)$, $x_1,x_2\in {\rm int}(X) $, $\eta x_1\le x_2\le \eta^{-1} x_1$ implies that $\eta \psi(t,x_1,y)\le \psi(t,x_2,y)\le \eta ^{-1}\psi(t,x_1,y)$, $\forall t\in \R, \forall y\in Y$; and
		\item[(II)] there exists $t_0>0$ and $y_0\in Y$ such that for any $\eta\in(0,1)$, $x_1,x_2\in {\rm int}(X) $, $\eta x_1\le x_2\le \eta^{-1} x_1$ implies that $\eta \psi(t_0,x_1,y_0)\ll \psi(t_0,x_2,y_0)\ll \eta^{-1} \psi(t_0,x_1,y_0)$.
	\end{itemize}
	If $H$ has a compact $\omega$-limit set $K_0\subset {\rm int}(X) \times Y$, then the natural projection $p: X\times Y\to Y$ is a flow isomorphism restricted on $H(t,\cdot,\cdot): K_0\times Y\to K_0$ and for every compact $\omega$-limit set $w(x,y) \subset {\rm int}(X)\times Y$, we have $w(x,y)=K_0$ and $\lim_{t\to\infty} \|\psi(t,x,y)- \psi(t,x^*,y)\|=0$, where $(x^*,y) = K_0\cap p^{-1}(y)$.
\end{theorem}

\subsection{Time-Dependent Linear Cooperative Systems}

We will leverage time-dependent linear systems and  $\delta$-matrices \cite{moreau2004stability}. 
Consider a time-dependent linear system
\begin{align}\label{tv}
\frac{dz}{dt} = B(t)z(t),
\end{align}
where $z(t)\in\R^{n}$ and $B(t)=[b_{ij}(t)]_{i,j=1}^{n}$ with
$b_{ij}(t)\ge 0$ for $i\ne j$ and $t\in\R$.
We will relate $B(t)$ to a time-dependent (attack-defense) graph $\G(B(t)) = (V,E(t),B(t))$ with node set $V=\{1,\cdots,n\}$, time-dependent arc set $E(t)$ and time-dependent weight matrix $B(t)$. A matrix $B^{\delta}(t) = [b^{\delta}_{ij}(t)]$ with
\begin{align*}
b_{ij}^{\delta}(t)=\begin{cases}b_{ij}(t)& b_{ij}(t)\ge \delta ~\&~ i\ne j\\
0& b_{ij}(t)<\delta~\&~i\ne j\\
b_{ii}(t)&i=j
\end{cases}
\end{align*} 
is called a {\em $\delta$-matrix }of $B(t)$ and its associated graph $\G(B^{\delta}(t))$ is called $\delta$-graph of $B(t)$. 

When $\{B(t)\}_{t\ge 0}$ is ergodic, meaning system \eqref{tv} is ergodic, let $M(B)$ denote its mean value (cf. Definition \ref{def-ergodic}). We will use the following lemma, which is a basic result regarding ergodic time-dependent linear system (\ref{tv}).
This lemma says that the fundamental solution matrix of system \eqref{tv} is positive when $\{B(t)\}_{t\in\R}$ is ergodic and the associated graph $\mathcal G(M(B))$ is strongly connected; its proof is deferred to Appendix \ref{appendix:lemma-linear-sys}.
\begin{lemma}\label{lemma-linear-sys}
	Consider system (\ref{tv}) with bounded and ergodic $\{B(t)\}_{t\ge 0}$. Let $M(B)$ denote the mean value of $\{B(t)\}_{t\ge 0}$ and $U(t,s)$ the fundamental solution matrix of system (\ref{tv}). Then, we have:
	\begin{itemize}
		\item[(i)] every element of $U(t,s)$ is nonnegative for any $t\ge s\ge 0$; and
		\item[(ii)] if $\G(M(B))$ is strongly connected, there exists $T>0$ such that for each $\Delta\ge T$, one can find $\epsilon>0$ (dependent on $\Delta$) such that every element of $U(s+\Delta,s)$ is greater than $\epsilon$ for any $s\ge 0$.
	\end{itemize}
\end{lemma}

\section{Model and Analysis}
\label{sec-new-results}

The preventive and reactive cyber defense dynamics model introduced in \cite{lin2018unified}, dubbed unified model with time-independent parameters, unifies two families of models into a single framework. The unified dynamics or model is proven to be {\em globally convergent} (i.e., always converging to some equilibrium) \cite{lin2018unified}. Now we present and analyze its extension to {\em unified model with time-dependent parameters}.

\subsection{The Unified Model with Time-Dependent Parameters}\label{sec-model}

The unified model with time-dependent parameters still describes the dynamics of the global cybersecurity state incurred by the interactions between two classes of cyber attacks (i.e., pull-based attacks and push-based attacks) and two classes of defenses (i.e., preventive defenses and reactive defenses) 
in a network. Intuitively, the attack-defense structure incurred by the attack-defense interactions at time $t$ can be described by a directed graph, denoted by $\tilde{G}(t)=(\tilde{V}(t),\tilde{E}(t))$, where $\tilde{V}(t)$ is the node (representing a computer) set and $\tilde{E}(t)$ is the arc set,
and $(u,v)\in \tilde{E}(t)$ means node $u$ can communicate with, and therefore can wage push-based attacks against, node $v$ at time $t$. 
As we will justify later, it suffices to consider a {\em time-independent} node set $V = \{1,\ldots,n\}$, namely $G(t)=(V,E(t))$ rather than $\tilde{G}(t)=(\tilde{V}(t),\tilde{E}(t))$, because the evolution of $\tilde{V}(t)$ can be ``encoded'' into the evolution of $E(t)$, leading to a simpler representation $G(t)=(V,E(t))$ of attack-defense structures.

The effect of attacks against preventive defenses is modeled as follows. Push-based attacks take place on attack-defense structures $G(t) = (V,E(t))$.
Denote by $A(t)$ the adjacency matrix of $G(t)$, where $a_{vu}(t)=1$ if $(u,v)\in E(t)$ and $a_{vu}(t) = 0$ otherwise. Let $\gamma_{vu}(t)\in[0,1]$ denote the probability that a push-based attack, which is waged by a compromised node $u\in V$ against a secure $v\in V$ over arc $(u,v)\in E(t)$ at time $t$, succeeds (i.e., causing $v$ to become compromised); that is, $1-\gamma_{vu}(t)$ represents the effectiveness of the preventive defense mechanism deployed at node $v$ and/or the arc $(u,v)$. Let $\Gamma(t)=[\gamma_{vu}(t)]_{n\times n}$ denote the probability matrix corresponding to the adjacency matrix $A(t)$. The effect of push-based attacks over $G(t)$ can be described by $(G(t),\Gamma(t))$. On the other hand, pull-based attacks can be described by using $\alpha_v(t)\in [0,1]$ to denote the probability that a secure node $v\in V$ becomes compromised at time $t$ because of pull-based attacks; that is, $1-\alpha_v(t)$ represents the effectiveness of preventive defense against pull-based attacks. 

For modeling the effect of reactive defenses against successful attacks, let $\beta_v(t)\in [0,1]$ denote the probability that a compromised node $v$ is detected and ``cleaned up'' (i.e., becoming secure) at time $t$; that is, $1-\beta_v(t)$ represents the ineffectiveness of the reactive defense. 

\begin{figure}[!htbp]
	\centering
	\begin{tikzpicture}[scale = 0.6]
	\tikzstyle{mystyle}=[ellipse, rounded corners,minimum width=3cm,minimum height=9mm,draw=black,fill=white,very thick,>=LaTeX]
	\node[mystyle,label=center: secure]        (sec)       at (-2,0)       {};
	\node[mystyle,label=center:  compromised]        (com)   at (4,0)        {};
	\draw[->,very thick] (sec) to [bend left = 20] node[above]{\small $g_{v}(\cdot,\cdot,\cdot)$} (com);
	\draw[->,very thick] (com) to [bend left = 20] node[below]{\small $h_{v}(\cdot,\cdot)$} (sec);
	\draw[->,very thick] (com) to [loop right,looseness=5] node[below right = 4mm and -2cm] {\small $1-h_{v}(\cdot,\cdot)$} (com);
	\draw[->,very thick] (sec) to [loop left, looseness=5] node[above left = 4mm and -2cm] {\small $1-g_{v}(\cdot,\cdot,\cdot)$} (sec);
	\end{tikzpicture}
	\caption{State transition diagram of node $v\in V$ at time $t$.}
	\label{fig-state-tran-diagram}
\end{figure}
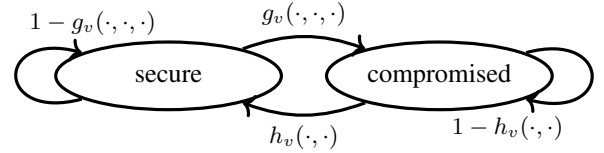

At any time $t$, a node $v\in V$ is in one of two cybersecurity states, {\em compromised} or {\em secure}. 
Let $i_v(t)$ and $s_v(t)$ respectively denote the probability that $v$ is in the {\em compromised} state and the {\em secure} state at time $t$. Let $i(t) = [i_1(t),\cdots,i_n(t)]$ and $s(t) = [s_1(t),\cdots,s_n(t)]$. 
Figure \ref{fig-state-tran-diagram} describes the state transition diagram of a node $v\in V$, where 
$h_v\left(i(t),\beta_v(t)\right): [0,1]^n \times [0,1] \to [0,1]$ is a function that outputs the probability that a compromised node $v$ becomes secure at time $t$ because of the deployed reactive defenses, and $g_v\left(i(t),\alpha_v(t),\Gamma(t)\right):  [0,1]^n \times [0,1]\times [0,1]^{n,n}\to [0,1]$ is a function that outputs the probability that a secure node $v$ becomes compromised at time $t$ because of the push-based and pull-based attacks that penetrate the deployed preventive defenses. Figure \ref{fig-state-tran-diagram} and the fact that $i_v(t)+s_v(t)=1$ holds for any $v\in V$ and any $t\ge 0$, the unified model with time-dependent parameters is described by the following dynamical system for each $v\in V$: 
\begin{align}
\nonumber
\frac{di_v(t)}{dt}= &-h_v\left(i(t),\beta_v(t)\right)\cdot i_v(t)\\ 
&+g_v\left(i(t),\alpha_v(t),\Gamma(t)\right)\cdot\left(1-i_v(t)\right). 
\label{eq:unified-generic-model}
\end{align}

The research objective is to analyze system or model \eqref{eq:unified-generic-model} with time-dependent parameters $\alpha_v(t)$,  $\beta_v(t)$, $\gamma_{vu}(t)$ and $G(t)=(V,E(t))$. This is a challenging task because among other things, the term $g_v\left(i(t),\alpha_v(t),\Gamma(t)\right)$ is highly nonlinear. In order to simplify presentation, we use $y(t)$ to denote the collection of parameters at time $t$, namely $y(t) = [\alpha(t),\beta(t),{\rm vec}(A(t))^{\top},{\rm vec}(\Gamma(t))^{\top}]$, where ${\rm vec}(matrix)$ is the vector representation of a matrix $matrix$ obtained by concatenating the columns; we use $f_v(i(t),y(t))$ to denote the right-hand side of system \eqref{eq:unified-generic-model}, namely
\begin{align}\nonumber
f_v(i(t),y(t))
&= -h_v\left(i(t),\beta_v(t)\right)\cdot i_v(t)\\\label{eq:f-general-form}
&+g_v\left(i(t),\alpha_v(t),\Gamma(t)\right)\cdot\left(1-i_v(t)\right);  
\end{align}
and we let $f(i,y) = [f_1(i,y),\cdots,f_n(i,y)]^{\top}$.

\medskip

\noindent{\bf On the generality of model \eqref{eq:unified-generic-model} in accommodating models studied in the literature}.
Model \eqref{eq:unified-generic-model} is a general framework because $h_v$ and $g_v$ can be instantiated in many ways. In particular, model \eqref{eq:unified-generic-model} degenerates to the unified model with time-independent parameters \cite{lin2018unified} by letting $G(t)$, $\gamma_{vu}(t)$, $\alpha_v(t)$ and $\beta_v(t)$ be time-independent. Moreover, the extended model also accommodates the aforementioned two models as special cases, namely the  $\prod$-model because a $\prod$-term instantiates $g_v$ as shown in Eq. (\ref{model-prod-sum}) and the $\sum$-model because a $\sum$-term instantiates $g_v$ as shown in Eq. (\ref{model-prod-sum}). 
\begin{align}\label{model-prod-sum}
&g_v\left(i(t),\alpha_v(t),\Gamma(t)\right) \\ \nonumber
= &\left\{ \begin{array}{ll} \displaystyle{1-(1-\alpha_v(t))\prod_{u\in N_v(t)} (1-\gamma_{vu}(t) i_u(t))}, & \text{$\prod$-model} \\ 
\displaystyle{\alpha_v(t) + \sum_{u\in N_v(t)} \gamma_{vu}(t) i_u(t)}, & \text{$\sum$-model}
\end{array}
\right.
\end{align}
Both models set $h_v\left(i(t),\beta_v(t)\right) = \beta_v(t)$. Both models are nonlinear, especially the $\prod$-model, which is highly nonlinear and thus makes the unified model with time-dependent parameters highly nonlinear. These two models differ in how they model the collective effect of push-based attacks waged by $v$'s neighbors against $v$, namely $g_v$.

\medskip

\noindent{\bf How can $G(t)=(V,E(t))$ encode $\tilde{G}(t)=(\tilde{V}(t),\tilde{E}(t))$}?
Given $\tilde{G}(t)=(\tilde{V}(t),\tilde{E}(t))$ over time $t$, we can set $V=\bigcup_t \tilde{V}(t)$, meaning that a node $u\in V -\tilde{V}(t)$ can be treated as an isolated, dummy node in $G(t)=(V,E(t))$ such that $(u,\cdot)\notin E(t)$ and $(\cdot,u)\notin E(t)$ and $\alpha_u(t)=0$.
This justifies why it suffices to consider $G(t)=(V,E(t))$.

\subsection{Some Properties of Functions $h_v$, $g_v$ and Parameters $y(t)$}

In order to make the model as widely applicable as possible, we need to make as few restrictions as possible on functions $h_v$ and $g_v$ in model \eqref{eq:unified-generic-model}. In order to facilitate analysis, we need functions $h_v$ and $g_v$ to have the following Properties \ref{pro-continuous}-\ref{property-ergodic}, which are both intuitive and natural.

\begin{property}
\label{pro-continuous}
For $\forall v$, $h_v\left(i,\beta_v(t)\right)$ and $g_v\left(i,\alpha_v(t),\Gamma(t)\right)$ have continuous first and second derivatives with respect to $i$.
\end{property}

Property \ref{pro-continuous} is the baseline for analytical treatment. 

\begin{property}
\label{pro-mono}
Consider $\forall v$ and $\forall u\neq v$, (i) $g_v(\mathbf{0},\alpha_v(t),\Gamma(t)) = \alpha_v(t)$, ${\partial g_v(i(t), \alpha_v(t),\Gamma(t))}/{\partial i_u}= 0$ when $\gamma_{vu}(t)=0$, and ${\partial h_v(i(t),\beta_v(t))}/{\partial i_u} = 0$; (ii) for any $\delta>0$, ${\partial g_v(i(t), \alpha_v(t),\Gamma(t))}/{\partial i_u} > \delta_1$ holds for some $\delta_1>0$ when $\gamma_{vu}(t)>\delta$.
\end{property}

Part (i) of Property \ref{pro-mono} abstracts the intuition that the probability node $v$ gets compromised at time $t$ is independent of the state of $u$ when $u$ cannot attack node $v$ or $(u,v)\notin E(t)$, and the probability node $u$ becomes secure at time $t$ is independent of the states of the other nodes at time $t$. Part (ii) of Property \ref{pro-mono} abstracts the intuition that the probability $v$ becomes compromised at time $t$ increases with the probability $u$ is compromised when $u$ can attack $v$  or $(u,v)\in E(t)$. 

\begin{property}
	\label{pro-subhomo}
	Consider $\forall v$, $g_v(i,\alpha_v(t),\Gamma(t))$ is subhomogeneous with respect to $i$ and $D_i(g_v(i,\beta_v(t))+h_v(i,\alpha_v(t),\Gamma(t))) = \left[\partial (g_v(i,\beta_v)+h_v(i,\alpha_v,\Gamma))/\partial i_u\right]_{u\in V} > \mathbf{0}$.
\end{property}

Property \ref{pro-subhomo} abstracts the intuition that there is always a nonzero probability for a compromised node $v$ to become secure because reactive defenses have a nonzero probability to succeed or $\beta_v(t)\geq 0$. 

In order to prove the global attractivity of model \eqref{eq:unified-generic-model} with time-dependent parameters, we need parameters $y(t)$ to satisfy the following property:

\begin{property}[ergodicity of $\{y(t)\}_{t\ge 0}$]
	\label{property-ergodic}
	$\{y(t)\}_{t\ge 0}$ is ergodic, where $y(t) = [\alpha(t),\beta(t),{\rm vec}(A(t))^{\top},{\rm vec}(\Gamma(t))^{\top}]$.
\end{property}

Note that Properties \ref{pro-continuous}-\ref{pro-subhomo} are naturally satisfied by the $\prod$-model and the $\sum$-model with time-dependent parameters, namely systems or models \eqref{model-prod-sum}. This is because Properties \ref{pro-continuous}-\ref{pro-subhomo} naturally extend their time-independent counterparts, which are satisfied by the $\prod$-model and the $\sum$-model with time-independent parameters.
We will show that Property \ref{property-ergodic} (i.e., ergodicity) is close to, if not, the necessary condition for the global attractivity of model \eqref{eq:unified-generic-model}, by presenting a numerical example to show that its violation disrupts the global attractivity. The validation of these four properties, while intuitive, is an orthogonal research problem to the present characterization study and will be investigated in the future.

\subsection{Model \eqref{eq:unified-generic-model} Is Strongly Subhomogeneous}

We start by showing that  model \eqref{eq:unified-generic-model} is strongly subhomogeneous under Properties \ref{pro-continuous} and \ref{pro-subhomo}.

\begin{theorem}
	\label{thm-pre-subhomogeneous}
	Model \eqref{eq:unified-generic-model} under Properties \ref{pro-continuous} and \ref{pro-subhomo} is strongly subhomogeneous in $[0,1]^n$. 
\end{theorem}

The proof of Theorem \ref{thm-pre-subhomogeneous} is similar to the proof of Lemma 4 in \cite{lin2018unified}, which considers the unified model with time-independent parameters. This is because the proof does not need to make any restrictions on the time-dependence of the parameters, meaning that the proof is equally applicable to both time-independent parameters (the setting of \cite{lin2018unified}) and our setting of time-dependent parameters.

\subsection{Special equilibrium $\mathbf{0}$ Is Globally Stable}

Theorem \ref{thm-alpha-0-stability} below shows that when there are no pull-based attacks, namely $\alpha_v(t)=0$, $\forall v\in V$, $\forall t$, the special equilibrium $\mathbf{0}$ of model \eqref{eq:unified-generic-model} is globally stable under a certain condition. Note that $\mathbf{0}$ is no equilibrium when some nodes are subject to pull-based attacks, namely that the average of $\alpha_v(t)$ over $t$ is positive for some $v\in V$. Proof of Theorem \ref{thm-alpha-0-stability} is deferred to Appendix \ref{appendix:thm-alpha-0-stability}.

\begin{theorem}[equilibrium $\mathbf{0}$ is globally stable under Properties \ref{pro-continuous}-\ref{pro-subhomo}]
	\label{thm-alpha-0-stability}
	Consider model (\ref{eq:unified-generic-model}) and $\alpha_v(t)=0$, $\forall v\in V$, $\forall t$. For any $y\in Y$ with $\mu(D_i f(\mathbf{0},y))<0$, equilibrium $\mathbf{0}$ is globally stable in $[0,1]^n$, namely that every trajectory $\psi(t,i_0,y)$ of model \eqref{eq:unified-generic-model} converges to $\mathbf{0}$ regardless of the initial value $i(0)\in[0,1]^n$.
\end{theorem}

\subsection{Model \eqref{eq:unified-generic-model} Is Globally Attractive}

\begin{theorem}[main result: the unified model with time-dependent parameters is globally attractive under Properties \ref{pro-continuous}-\ref{property-ergodic}] 
	\label{thm-general}
	Model \eqref{eq:unified-generic-model} is almost surely globally attractive in $[0,1]^n\bo$, meaning that for almost every $y\in Y$, there exists a unique trajectory $\psi(t,i^*_y,y)\in[0,1]^n$ such that $\lim_{t\to\infty}\|\psi(t,i_0,y)-\psi(t,i^*_y,y)\|=0$ holds for any $i_0\in[0,1]^n\bo$. Moreover, $\forall v\in V$, $[\psi(t,i^*_y,y)]_v\neq 0$ as long as node $v$ is subject to pull-based attacks, meaning that the average of $\alpha_v(t)$ over time is positive.
\end{theorem}

Note that equilibrium $\mathbf{0}$ can be seen as a special case of a trajectory, but we separate its treatment (Theorem \ref{thm-alpha-0-stability}) from the treatment of the global attractivity in $[0,1]^n\bo$ (Theorem \ref{thm-general}) because the former, as a special case, can be proven without requiring the parameters $\{y(t)\}_{t\geq 0}$ to be ergodic (Property \ref{property-ergodic}).

The proof of Theorem \ref{thm-general} is deferred to Appendix \ref{appendix:thm-general}. In order to prove Theorem \ref{thm-general},  we need the following Lemmas \ref{lemma-persistent}--\ref{lemma-disconnect}, whose proofs are respectively deferred to Appendices \ref{appendix:lemma-persistent}--\ref{appendix:lemma-disconnect}. 
In the following lemmas, we let $M(\Gamma)$ denote the mean value of the ergodic process $\{\Gamma(t)\}_{t\ge 0}$ and call the weighted graph $\mathcal G(M(\Gamma))$ the {\em mean attack-defense structure} of $(G(t),\Gamma(t))$. 
Lemmas \ref{lemma-persistent}--\ref{lemma-disconnect} cope with different settings of the parameters (following the ``divide and conquer'' strategy): there are pushed-based and pull-based attacks vs. there are only pushed-based attacks (i.e., no nodes are subject to pull-based attacks); the mean attack-defense structure is strongly connected vs. it is not strongly connected.
Lemmas \ref{lemma-persistent}-\ref{lemma-alpha-0-ergodic} deal with the case that there are no pull-based attacks, namely $\alpha_v(t) =0$, $\forall v\in V$, $\forall t$;
Lemma \ref{lemma-alpha>0} deals with the case that there is at least some node $v\in V$ that is subject to pull-based attack, namely that these is at least one $v\in V$ such that the average of $\alpha_v(t)$'s over time $t$ is positive;
Lemma \ref{lemma-disconnect} deals with the case that the mean attack-defense structure $\mathcal G(M(\Gamma))$ has two strongly connected components, where the {\em mean} is averaged over time.
Note that it suffices to consider two strongly connected components because multiple strongly connected components can be treated in the same fashion and any attack-defense structure can be partitioned into strongly connected components. 

\subsubsection{Global attractivity when there are no pull-based attacks}
Theorem \ref{thm-pre-global-attractivity} says that model \eqref{eq:unified-generic-model} is globally attractive when the dynamics is {\em uniformly persistent}  (Definition \ref{def-persistence}) and satisfies conditions (I) and (II) of Theorem \ref{thm-pre-global-attractivity}. 
We first prove that model \eqref{eq:unified-generic-model} is {\em uniformly persistent} when $\mu(D_i f(\mathbf{0},y))>0$ and the mean attack-defense structure $\mathcal G(M(\Gamma))$ is strongly connected.

\begin{lemma}[model \eqref{eq:unified-generic-model} is uniformly persistent when there are no pull-based attacks, under Properties \ref{pro-continuous}-\ref{property-ergodic}]
	\label{lemma-persistent}
	If $\mu(D_i f(\mathbf{0},y))>0$ and the mean attack-defense structure $\mathcal G(M(\Gamma))$ is strongly connected, trajectories of model \eqref{eq:unified-generic-model} with nonzero initial values are uniformly persistent, namely $\lim\inf_{t\to+\infty} \psi(t,i_0,y)\gg\mathbf{0}$, $\forall i_0\in [0,1]^n\bo$ and $\forall y\in Y$.
\end{lemma}

Lemma \ref{lemma-alpha-0-ergodic} below shows that model \eqref{eq:unified-generic-model} is {\em globally attractive} (Definition \ref{def-global-attractive}) when Properties \ref{pro-continuous}-\ref{property-ergodic} hold and the mean attack-defense structure is strongly connected. 

\begin{lemma}[model \eqref{eq:unified-generic-model} is globally attractive when there are no pull-based attacks under Properties \ref{pro-continuous}-\ref{property-ergodic}]
	\label{lemma-alpha-0-ergodic}
	Suppose the mean attack-defense structure $\mathcal G(M(\Gamma))$ is strongly connected. 
	\begin{itemize}
		\item[(i)]If $\mu(D_i f(\mathbf{0},y))>0$, then there exists, for almost every $y\in Y$, a positive trajectory $\psi(t,i^*_y,y)\in (0,1]^n$ that is globally attractive in $[0,1]^n\bo$.
		\item[(ii)] If $\mu(D_i f(\mathbf{0},y)) = 0$, then there exists, for almost every $y\in Y$, a trajectory $\psi(t,i^*_y,y)\in [0,1]^n$ that is globally attractive in $[0,1]^n\bo$. Moreover, if $\psi(t,i^*_{y_0},y_0)\in (0,1]^n$ for some $y_0\in Y$, then for almost every  $y\in Y$, its associated globally attractive trajectory $\psi(t,i^*_y,y)\in (0,1]^n$.
		\item[(iii)] If $\mu(D_i f(\mathbf{0},y)) < 0$, the origin  $\mathbf{0}$ is almost surely globally attractive, i.e., $\lim_{t\to\infty}\psi(t,i_0,y) = \mathbf{0}$ holds for any $i_0\in [0,1]^n$ and almost every $y\in Y$.
	\end{itemize}
\end{lemma}

\subsubsection{Global attractivity when there are pull-based attacks}

From the ergodicity and non-negativity of the $\alpha_v(t)$'s, it follows that the mean value of $\{\alpha_v(t)\}_{t\ge 0}$ over $t$, denoted by $M(\alpha_v)$, is zero, namely $M(\alpha_v)=0$, if and only if $\alpha_v(t)=0$ almost surely. Therefore, when $\alpha_v(t)$ is ergodic, node $v$ is subject to pull-based attacks if and only if the mean value $M(\alpha_v)>0$. 
In what follows, we will use $\alpha_v(t)>0$ to denote that the mean value $M(\alpha_v)$ is positive. Let $V_{\alpha> 0} = \{v: M(\alpha_v)>0\}$ denote the set of nodes that are subject to pull-based attacks. The global attractivity of model (\ref{eq:unified-generic-model}) when $V_{\alpha>0}\neq \emptyset$ is established by the following lemma, while noting that $\mathbf{0}$ is no longer an equilibrium of model (\ref{eq:unified-generic-model}) when some nodes are subject to pull-based attacks. 

\begin{lemma}[model  (\ref{eq:unified-generic-model}) is globally attractive when there are pull-based attacks under Properties \ref{pro-continuous}-\ref{property-ergodic}]
	\label{lemma-alpha>0}
	If $V_{\alpha > 0}\neq \emptyset$ and the mean attack-defense structure $\mathcal G(M(\Gamma))$ is strongly connected, then there exists, for almost every $y\in Y$, a unique positive trajectory $\psi(t,i^*_y,y)$ that is globally attractive in $[0,1]^n$, namely $\lim_{t\to\infty}\|\psi(t,i_0,y)-\psi(t,i^*_y,y)\|=0$ holds $\forall i_0\in[0,1]^n$.
\end{lemma}

\subsubsection{Global attractivity when the mean attack-defense structure is not strongly connected}
Lemmas \ref{lemma-alpha-0-ergodic}-\ref{lemma-alpha>0} show model (\ref{eq:unified-generic-model}) is globally attractive when the mean attack-defense structure $\mathcal G(M(\Gamma))$ is strongly connected. Now we consider the case that  $\mathcal G(M(\Gamma))$ is not strongly connected, but can be divided into two strongly connected components (SCCs), denoted by $SCC_1$ and $SCC_2$. When there are no links between $SCC_1$ and $SCC_2$, the global attractivity for each $SCC_k$, $k=1,2$ is obtained from Lemma \ref{lemma-alpha-0-ergodic} and Lemma \ref{lemma-alpha>0} directly by treating each SCC as an attack-defense structure. Therefore, we only need to consider the case that there exist links between these two strongly connected components of $\mathcal G (M(\Gamma))$. 

From the Perron-Frobenius theory \cite{berman1994nonnegative}, $M(\Gamma)$ can be written in the lower-triangular block form, 
$$
M(\Gamma) = P\left[\begin{array}{ll}
\tilde{\Gamma}_{11}& 0\\
\tilde{\Gamma}_{21}& \tilde{\Gamma}_{22}\end{array}\right]P^{\top}
$$
where $P$ is a permutation matrix, $\tilde{\Gamma}_{kk}$ is irreducible and corresponds to $SCC_{k}$ for $k=1,2$, and $\tilde{\Gamma}_{21}\neq 0$. It follows from Property \ref{pro-mono} that for any $v\neq u$, ${\partial f_v(\mathbf{0},y)}/{\partial i_u} = {\partial g_v(\mathbf{0},\alpha_v,\Gamma)}/{\partial i_u}=0$ holds if $\gamma_{vu}(t)=0$, implying that the Jacobian matrix $D_i f(\mathbf{0},y)$ can be written as
\begin{align*}
D_i f(\mathbf{0},y)  = P\left[\begin{array}{ll}
D f_{11}& 0\\
D f_{21}& D f_{22}\end{array}\right]P^{\top},
\end{align*}
where $D f_{kk}$ corresponds to $SCC_{k}$ for $k=1,2$. Let $\V_{SCC_{k}}$ denote the set of nodes in $SCC_k$ and $|SCC_k|$ denote the number of nodes in $SCC_k$, where $k=1,2$. 
The following lemma considers the general case of $\alpha_v(t)\ge 0$, $\forall v\in V$.
 
\begin{lemma}[model (\ref{eq:unified-generic-model}) is globally attractive when the mean attack-defense structure has two strongly connected components, under Properties \ref{pro-continuous}-\ref{property-ergodic}] 
	\label{lemma-disconnect}
	Suppose $\mathcal G(M(\Gamma))$ consists of two strongly connected components $SCC_1$ and $SCC_2$ such that (without loss of generality) at least one node in $SCC_1$ has a path to a node in $SCC_2$. For $SCC_1$, there exists a globally attractive trajectory $\phi(t,i_{SSC_1}^*,y)\in[0,1]^{|SCC_1|}$ such that
	\begin{itemize}
		\item[(i)] if $\phi(t,i_{SCC_1}^*,y) = \mathbf{0}_{|SCC_1|}$, $\V_{SCC_2}\cap V_{\alpha > 0} = \emptyset$ and $\mu(Df_{22})<0$, the origin $\mathbf{0}_n$ is globally attractive for $\V_{SCC_1}\cup \V_{SCC_2}$, meaning that every trajectory of model (\ref{eq:unified-generic-model}) converges to $\mathbf{0}_n$ regardless of the initial values;
		\item[(ii)] otherwise, there is a trajectory $\psi(t,i^*,y)$ that is globally attractive in $[0,1]^n\backslash\{\mathbf{0}_n\}$, where $[\psi(t,i^*,y)]_{SCC_1} = \phi(t,i_{SCC_1}^*,y)$. Moreover, if $\V_{SCC_2}\cap     V_{\alpha > 0} \neq \emptyset$, $[\psi(t,i^*,y)]_{SCC_2}\in (0,1]^{|SCC_2|}$ is globally attractive in $[0,1]^{|SCC_2|}$.
	\end{itemize}
\end{lemma}

\subsection{Stronger Assumptions Leading to Stronger Results}
Now we show that stronger results can be obtained when the parameters exhibit stronger properties than ergodicity, such as {\em (almost) periodic}
(Definition \ref{def-almost-periodic}). 

\begin{theorem}[model \eqref{eq:unified-generic-model} is globally attractive and almost periodic when its parameters are almost periodic]
	\label{thm-almost-periodic-global}
	Consider model (\ref{eq:unified-generic-model}) under Properties \ref{pro-continuous}-\ref{pro-subhomo} and $y(t)$ is almost periodic (i.e., a property stronger than Property \ref{property-ergodic}): (i) there exists an almost periodic trajectory $\psi(t,i^*,y)\in[0,1]^n$ that is globally attractive in $[0,1]^n\bo$, namely that $\lim_{t\to\infty}\|\psi(t,i_0,y)-\psi(t,i^*,y)\|=0$ holds for any $i_0\in[0,1]^n\bo$;
	(ii) for any $\epsilon>0$, there exists $M>0$ so that the $\epsilon$-translation numbers of $y(t)$ are  $M\epsilon$-translation numbers of the globally attractive trajectory $\psi(t,i^*,y)$.
\end{theorem}

Proof of Theorem \ref{thm-almost-periodic-global} is deferred to Appendix \ref{appendix:thm-almost-periodic-global}. Since periodic functions are almost periodic by setting $\epsilon = 0$ in Definition \ref{def-almost-periodic}, we obtain the following corollary.

\begin{corollary}[model \eqref{eq:unified-generic-model} is globally attractive and periodic when its parameters are periodic]
	\label{coro-period}
	Consider model (\ref{eq:unified-generic-model}) under Properties \ref{pro-continuous}-\ref{pro-subhomo}. Suppose $y(t)$ is periodic with period $w$, then there exists an $w$-periodic trajectory $\psi(t,i^*,y)$ of model (\ref{eq:unified-generic-model}) that is globally attractive in $[0,1]^n\bo$, namely that $\lim_{t\to\infty}\|\psi(t,i_0,y)-\psi(t,i^*,y)\|=0$ holds for any $i_0\in[0,1]^n\bo$.
\end{corollary}

\subsection{Bounding the Globally Attractive Positive Trajectory}

When given all of the parameters, one can numerically compute the globally attractive trajectory. In practice, we may not know the values of all parameters (i.e., the matter of partial vs. full information). 
In this case, a useful alternative is to bound the globally attractive trajectory because, for example, we can treat the upper bound as the worst case scenario in cyber defense decision-making.
In order to derive such bounds, we need $h_v$ and $g_v$ to have the following intuitive property:

\begin{property}[properties of $h_v$ and $g_v$ needed for bounding the globally attractive trajectory]
	\label{pro-mono-2}
	Consider $g_v(i,\alpha_v,\Gamma)$ and $h_v(i,\beta_v)$, it holds that
	\begin{itemize}
		\item[(i)] $\partial g_{v}/\partial \alpha_v\ge 0$ and $\partial g_{v}/\partial \gamma_{vu}\ge 0$ for $\forall u,v\in V$, which reflects the intuition that the probability node $v$ getting compromised increases with the pull-based and push-based attack capabilities; and 
		\item[(ii)] $\partial h_{v}/\partial \beta_v\ge 0$ for $\forall v\in V$, which reflects the intuition that when everything else is fixed, the probability a compromised node $v$ becoming secure increases with the reactive defense capability.
\end{itemize}
\end{property}

To simplify the presentation, in the rest of this subsection we will use the following notations. For any $u,v \in V$, we set
{\small
\begin{eqnarray*}
\underline{\alpha}_v = \min_{t\ge 0} \alpha_v(t),~
&\overline{\alpha}_v = \displaystyle \max_{t\ge 0} \alpha_v(t),\\
\underline{\beta}_v = \min_{t\ge 0} \beta_v(t),~
&\overline{\beta}_v = \displaystyle \max_{t\ge 0} \beta_v(t),\\
\underline{\gamma}_{vu} = \min_{t\ge 0} \gamma_{vu}(t),~
&\overline{\gamma}_{vu} = \displaystyle \max_{t\ge 0} \gamma_{vu}(t),\\
\underline{\Gamma} = [\underline{\gamma}_{vu}]_{v,u\in V},
&~\overline{\Gamma} = [\overline{\gamma}_{vu}]_{v,u\in V}, \\
i^{\min}_v= \inf_{t\ge 0}\{i_v(t)\},
&i^{\max}_v = \displaystyle \sup_{t\ge 0}\{i_v(t)\},\\
i_{\min} = [i^{\min}_1,\cdots,i^{\min}_n], 
&i_{\max} = [i^{\max}_1,\cdots,i^{\max}_n],\\
\underline{h}_v(\beta_v) = \min_{i\in[i_{\min},i_{\max}]}h_v(i,\beta_v),&
\overline{h}_v(\beta_v) = \max\limits_{i\in[i_{\min},i_{\max}]}h_v(i,\beta_v).
\end{eqnarray*}
}

The basic idea is to leverage $i_{\min}$ and $i_{\max}$ to derive the bounds.
When $i_{\min}$ and $i_{\max}$ are unknown, we can set $i_{\min}=\mathbf{0}$ and $i_{\max}=\mathbf{1}$.
On the other hand, the properties $\partial g_v/\partial i_u\ge 0$, $\partial g_v/\partial \alpha_v\ge 0$ and $\partial g_v/\partial \gamma_{vu}\ge 0$ lead to that $g_v$ is monotone with respect to $i,\alpha_v$ and $\gamma_{vu}$, which implies 
$$
g_v(i_{\min},\underline{\alpha}_v,\underline{\Gamma})\le g_v(i(t),\alpha_v,\Gamma)\le g_v(i_{\max},\overline{\alpha}_v,\overline{\Gamma}).
$$
Then, it follows from $\partial h_{v}/\partial \beta_v\ge 0$ that 
$$
\underline{h}_v(\underline{\beta}_v) \le  h_v(i,\beta_v) \le \overline{h}_v(\overline{\beta}_v).
$$

The following theorem can bound the dynamics when only knowing the lower and upper bounds of parameters $\alpha_v(t)$, $\beta_v(t)$, and $\gamma_{uv}(t)$, $\forall u,v\in V$, $\forall t$, while the $G(t)$'s are given. Proof of Theorem \ref{thm-bounds} is deferred to Appendix \ref{appendix:thm-bounds}.

\begin{theorem}[bounding the globally attractive trajectory]
	\label{thm-bounds}
	Let $i_v(t)$ be the solution to model (\ref{eq:unified-generic-model}) and $\overline{i}_v(t)$ and $\underline{i}_v(t)$ be the upper and lower bounds of $i_v(t)$. Then under Property \ref{pro-mono-2}, we have $\underline{i}_v(t)\le i_v(t)\le \overline{i}_v(t)$ for $\forall v\in V, t\in \R_{\ge 0}$, where 
	\begin{align}\label{lower-bound}
	\underline{i}_v(t)= \exp(-\underline{\mathcal A}_vt)\left[\underline{i}_v(0)-\frac{\underline{\mathcal B}_v}{\underline{\mathcal A}_v}\right]+\frac{\underline{\mathcal B}_v}{\underline{\mathcal A}_v}
	\end{align}
	\begin{align}\label{upper-bound}
	\overline{i}_v(t)= \exp(-\overline{\mathcal A}_vt)\left[\overline{i}_v(0)-\frac{\overline{\mathcal B}_v}{\overline{\mathcal A}_v}\right]+\frac{\overline{\mathcal B}_v}{\overline{\mathcal A}_v}
	\end{align}
	with $\underline{\mathcal A}_v$, $\overline{\mathcal A}_v$, $\underline{\mathcal B}_v$, and $\overline{\mathcal B}_v$ satisfying
	\begin{align*}
	&\left\{\begin{array}{l}
	\underline{\mathcal A}_v =\overline{h}_v(\overline{\beta}_v) + g_v(i_{\min},\underline{\alpha}_v,\underline{\Gamma}),\\
	\underline{\mathcal B}_v =g_v(i_{\min},\underline{\alpha}_v,\underline{\Gamma}),\\
	\overline{\mathcal A}_v =\underline{h}_v(\underline{\beta}_v)+g_v(i_{\max},\overline{\alpha}_v,\overline{\Gamma}),\\
	\overline{\mathcal B}_v =g_v(i_{\max},\overline{\alpha}_v,\overline{\Gamma}).
	\end{array}\right.
	\end{align*}
\end{theorem}

\subsection{Relationship between the Literature Results and Ours}

Now we show that our results supersede the state-of-the-art results because they are equivalent to corollaries of our results.

\begin{corollary}[corollary of our Theorem \ref{thm-general} and Lemmas \ref{lemma-alpha-0-ergodic}-\ref{lemma-disconnect} is equivalent to Theorem 3 in \cite{lin2018unified}]
\label{corollary-ours-supersedes-lins}
Consider model (\ref{eq:unified-generic-model}) under Properties \ref{pro-continuous}-\ref{pro-subhomo}. Suppose parameters $\alpha_v(t)$, $\beta_v(t)$, $\gamma_{vu}(t)$ are time-independent for all $u,v\in V$. Suppose the attack-defense structure $G(t)$ is also time-independent and contains $K$ strongly connected components, denoted by $SCC_k$ for $k=1,\cdots,K$, and the Jacobian matrix $D_i f(\mathbf{0},y)$ has the following Perron-Frobenius form
{
\small
\begin{align}\label{Df-PF-form-K}
&D_i f(\mathbf{0},y) =
P\left[\begin{array}{llll}
D f_{11}& 0& \cdots& 0\\
D f_{21}& D f_{22}& \cdots& 0\\
\vdots& \vdots& \ddots& \vdots\\
D f_{K1}& D f_{K2}& \cdots& Df_{KK}\\
\end{array}\right]P^{\top}
\end{align}
}
where $P$ is a permutation matrix, and $D f_{kk}$ corresponds to $SCC_{k}$ for $k=1,\cdots,K$. Let $R_k$ be the indices of the $SCC$'s that have links pointing to $SCC_k$. Then, for each $SCC_k$, we have:
\begin{itemize}
\item[(i)]$\mathbf{0}_{|SCC_k|}$ is globally asymptotically stable in $[0,1]^n$ if one of the following conditions holds:
\begin{itemize}
\item $\V_{SCC_k}\cap V_{\alpha>0}= \emptyset$, $\mu(Df_{kk})\le 0$ and $SCC_{R_k}= \emptyset$;
\item $\V_{SCC_k}\cap V_{\alpha>0}= \emptyset$, $\mu(Df_{kk})\le 0$, $SCC_{R_k}\neq \emptyset$ and $\mathbf{0}_{|SCC_r|}$ is globally asymptotically stable for every $r\in {R_k}$.
\end{itemize}
\item[(ii)] $SCC_k$ has a unique positive equilibrium that is globally asymptotically stable in $[0,1]^{|SCC_k|}$ if $\V_{SCC_k}\cap V_{\alpha>0}\neq \emptyset$.
\item[(iii)]  $SCC_k$ has a unique positive equilibrium that is globally asymptotically stable in $[0,1]^{|SCC_k|}\bo$ if one of the following conditions holds:
\begin{itemize}
\item $SCC_k\cap\V_{\alpha>0}= \emptyset$ and $\mu(Df_{kk})>0$;
\item $SCC_k\cap\V_{\alpha>0}= \emptyset$, $SCC_{R_k}\neq \emptyset$ and $\mathbf{0}_{SCC_r}$ is not globally asymptotically stable for every $r\in {R_k}$.
\end{itemize}
\end{itemize}
\end{corollary}

The state-of-the-art result of the $\sum$-model with time-dependent $G(t)$ is given in  \cite{bokharaie2010spread,pare2018epidemic}, and the state-of-the-art result of the $\prod$-model with time-dependent $G(t)$ is given in \cite{XuTAAS2014,zhang2017spectral,prakash2010virus,sanatkar2016epidemic}. However, these studies only investigate the stability of the equilibrium $\mathbf{0}$. Among these studies,  \cite{zhang2017spectral,prakash2010virus,sanatkar2016epidemic,bokharaie2010spread} consider the discrete-time model and show that the equilibrium $\mathbf{0}$ is stable if the {\em joint spectral radius} of the set of system matrices is smaller than 1. For a discrete-time system $z(t+1) = C(t) z(t)$, $t\in\mathbb{N}$, let $\mu_d(C(t)) = \lim_{n\to\infty} 1/n \log(\|C(n)\cdots C(1)C(0)\|)$ be the associated maximum Lyapunov exponent (MLE) and $\rho(C(t))$ be the joint spectral radius given in \cite{zhang2017spectral,prakash2010virus,sanatkar2016epidemic,bokharaie2010spread}. In fact, the joint spectral radius can be seen as the discrete-time version of the MLE; i.e.,  $\rho(C(t))<1$ is equivalent to $\mu_d(C(t))<0$
and thus their result is a special case of our Theorem \ref{thm-alpha-0-stability}. 
\begin{corollary}[corollary of our Theorem \ref{thm-alpha-0-stability} is equivalent to Theorem 1 in \cite{zhang2017spectral}, Theorem 2 in \cite{prakash2010virus}, Theorem 1 in \cite{sanatkar2016epidemic} and Theorem 2.1 in \cite{bokharaie2010spread}]
	\label{corollary-ours-supersedes-zhang}
	Consider the discrete-time version of model \eqref{eq:unified-generic-model}. Suppose $\alpha_v(t) = 0$ and $\gamma_{vu}(t) = \gamma(t) $, $\forall u,v\in V$, $\forall t$. Let $B(t)={\rm diag}(\beta_1(t),\dots,\beta_n(t))$. If $\rho((I-B(t))+\gamma(t) A(t))<1$, the dynamics of the discrete-time model globally converges to equilibrium $\mathbf{0}$.
\end{corollary}

The $\sum$-model with time-dependent parameters is studied in \cite{pare2018epidemic} and shown to converge to the equilibrium $\mathbf{0}$ when the time-dependent parameters satisfy some specific conditions. For symmetric matrices $C(t),t\ge 0$, we have $\mu(C(t))\le \sup_{t\ge 0}\lambda_1(C(t))$, meaning that the following corollary of our Theorem \ref{thm-alpha-0-stability} is equivalent to the result of \cite{pare2018epidemic}.

\begin{corollary}[corollary of our Theorem \ref{thm-alpha-0-stability} is equivalent to Theorem 1 in \cite{pare2018epidemic}]
\label{corollary-ours-supersedes-pares}
Consider model \eqref{eq:unified-generic-model} while instantiating its term $g_v(i(t),\alpha_v(t),\Gamma(t))$ as the $\sum$-model with time-dependent parameters. Suppose $\alpha_v(t) = 0$ and $\Gamma(t)$ being symmetric for  $\forall v\in V$, $\forall t$. Let $B(t)={\rm diag}(\beta_1(t),\dots,\beta_n(t))$. If $\sup_{t\ge 0}\lambda_1(\Gamma(t)-B(t))<0$, the dynamics of \eqref{eq:unified-generic-model} globally converges to equilibrium $\mathbf{0}$.
\end{corollary}

The $\prod$-model with time-dependent parameters is considered in \cite{XuTAAS2014}, which shows when the dynamics converges to the equilibrium $\mathbf{0}$. The following corollary of our Theorems \ref{thm-alpha-0-stability} and \ref{thm-general} is equivalent to the result of \cite{XuTAAS2014}.

\begin{corollary}[corollary of our Theorems \ref{thm-alpha-0-stability} and \ref{thm-general} is equivalent to Theorem 1 in \cite{XuTAAS2014}]
	\label{corollary-ours-supersedes-adaptive}
	Consider model \eqref{eq:unified-generic-model} while instantiating its term $g_v(i(t),\alpha_v(t),\Gamma(t))$ as the $\prod$-model with time-dependent parameters. Suppose $\alpha_v(t) = 0$ and $\gamma_{vu}(t) = \gamma(t) $, $\forall u,v\in V$, $\forall t$. Let $B(t)={\rm diag}(\beta_1(t),\dots,\beta_n(t))$. If $\mu(\gamma(t)A(t)-B(t))<0$, model (\ref{eq:unified-generic-model}) in this special case converges to equilibrium $\mathbf{0}$; if $\mu(\gamma(t)A(t)-B(t))>0$ and the linear system $dz(t)/dt = (\gamma(t)A(t)-B(t)) z(t)$ is ergodic, equilibrium $\mathbf{0}$ is unstable.
\end{corollary}

\subsection{Systematizing Knowledge}

\begin{figure}[htbp]
    \centering
        \includegraphics[clip, trim= 1cm 0.6cm 9cm 0.9cm, width=0.5\textwidth]{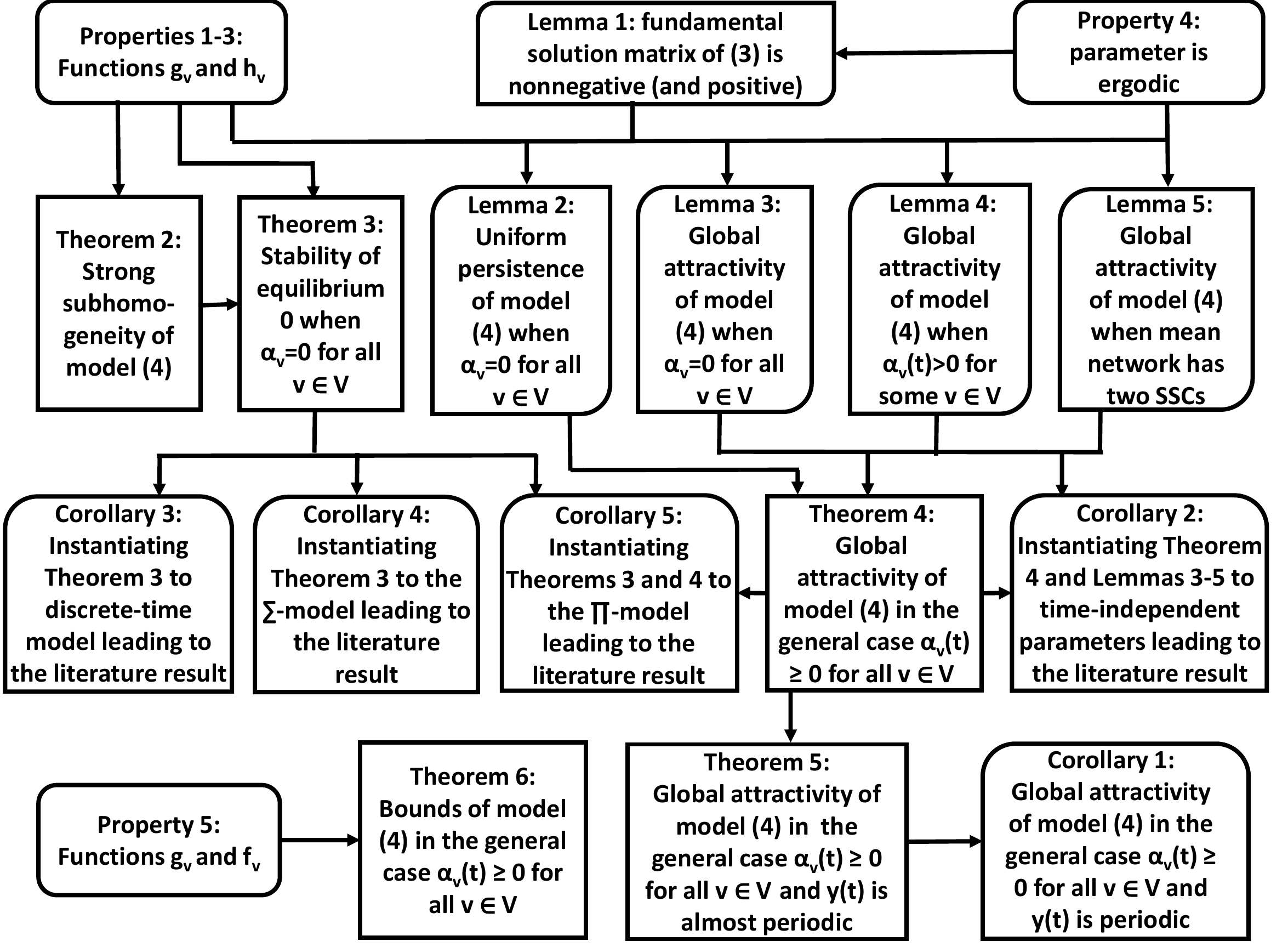}
        \caption{Relationship between the mathematical properties, lemmas, theorem, and corollaries presented in the paper.}
        \label{fig:relationship}
\end{figure}

We use Figure \ref{fig:relationship} to systematize the relationship between the properties, lemmas, theorems, and corollaries (i.e., their equivalent literature results). 

\section{Numerical examples}
\label{sec-numerical}

We use numerical results to confirm our analytic results.
In our experiments, we use the Euler method for the numerical simulation of model \eqref{eq:unified-generic-model} by setting the iteration step as 0.05.
In order to succinctly present the experimental result, we plot the dynamics of $\langle i(t)\rangle=\sum_{v}i_v(t)/|V|$, which is the fraction of compromised nodes at time $t$. In our experiment, we set the initial fraction of compromised nodes as $\langle i(0)\rangle \in \{0.25,0.5,0.75\}$, meaning that $25\%$, $50\%$, and $75\%$ randomly chosen nodes are in the compromised state at time $t=0$. 

\subsection{Confirming Global Attractivity}
In our experiments, we use the Gnutella05 peer-to-peer network {http://snap.stanford.edu/data/} as the initial attack-defense structure $G(0)$, where $|V| = 8846$ nodes and $|E(0)|= 31839$ arcs. In order to generate $G(t)$ for $t>0$, we randomly add or delete 2\% of the arcs of the attack-defense structure after every 10 simulation time units. In our experiments, we consider the $\sum$-model with time-dependent parameters $h_v(i,\beta_v) = \beta_v(t)$ and $g_v(i,\alpha_v,\Gamma)$ given in model \eqref{model-prod-sum}, which satisfies Properties \ref{pro-continuous}-\ref{pro-subhomo} as required by the main results (i.e., Theorem \ref{thm-general}).

In the first experiment, we set $\alpha_v(t)=0$, $\forall v$, $\forall t$, 
and use the following parameter sets
\begin{itemize}
	\item (p1): $\beta_v(t)= 0.1\sin(t) + 0.1 \sin(\sqrt{2} t)+0.5$, $\forall v\in V$, and $\gamma_{uv}(t)=0.05\sin(\pi t/5)+0.1$, $\forall (u,v)\in E(t)$;
	\item (p2): $\beta_v(t)= 0.1\sin(t) + 0.1 \sin(\sqrt{2} t)+0.4$, $\forall v\in V$, and $\gamma_{uv}(t)=0.05\sin(\pi t/5)+0.1$, $\forall (u,v)\in E(t)$;
	\item (p3): $\beta_v(t)= 0.1\sin(t) + 0.1 \sin(\sqrt{2} t)+0.1$, $\forall v\in V$, and $\gamma_{uv}(t)=0.05\sin(\pi t/5)+0.1$, $\forall (u,v)\in E(t)$;
	\item (p4): $\gamma_{uv}(t)=0.3\sin(\pi t/5)+0.7$, $\forall u,v$; multiple $\beta_v(t)$'s that will be specified below.
\end{itemize}
Note that parameter sets (p1), (p2) and (p3) correspond to the cases $\mu(D_{i}f(\mathbf{0},y(t)))<0$, $\mu(D_{i}f(\mathbf{0},y(t)))=0$ and $\mu(D_{i}f(\mathbf{0},y(t)))>0$, respectively. 

\begin{figure}[!htbp]
	\centering
	\begin{subfigure}[b]{0.23\textwidth}
		\centering
		\includegraphics[width=\textwidth]{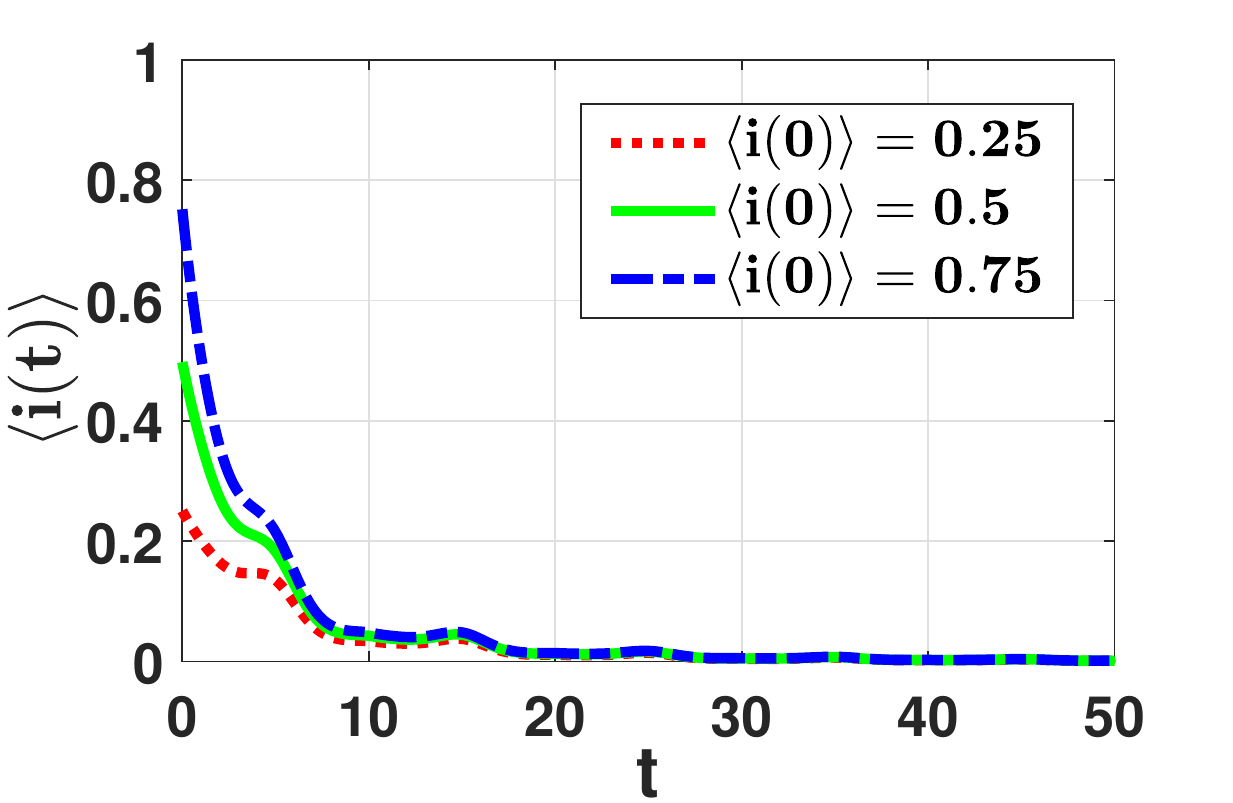}
		\caption{Parameter set (p1) }
	\end{subfigure}
	\begin{subfigure}[b]{0.23\textwidth}
		\centering
		\includegraphics[width=\textwidth]{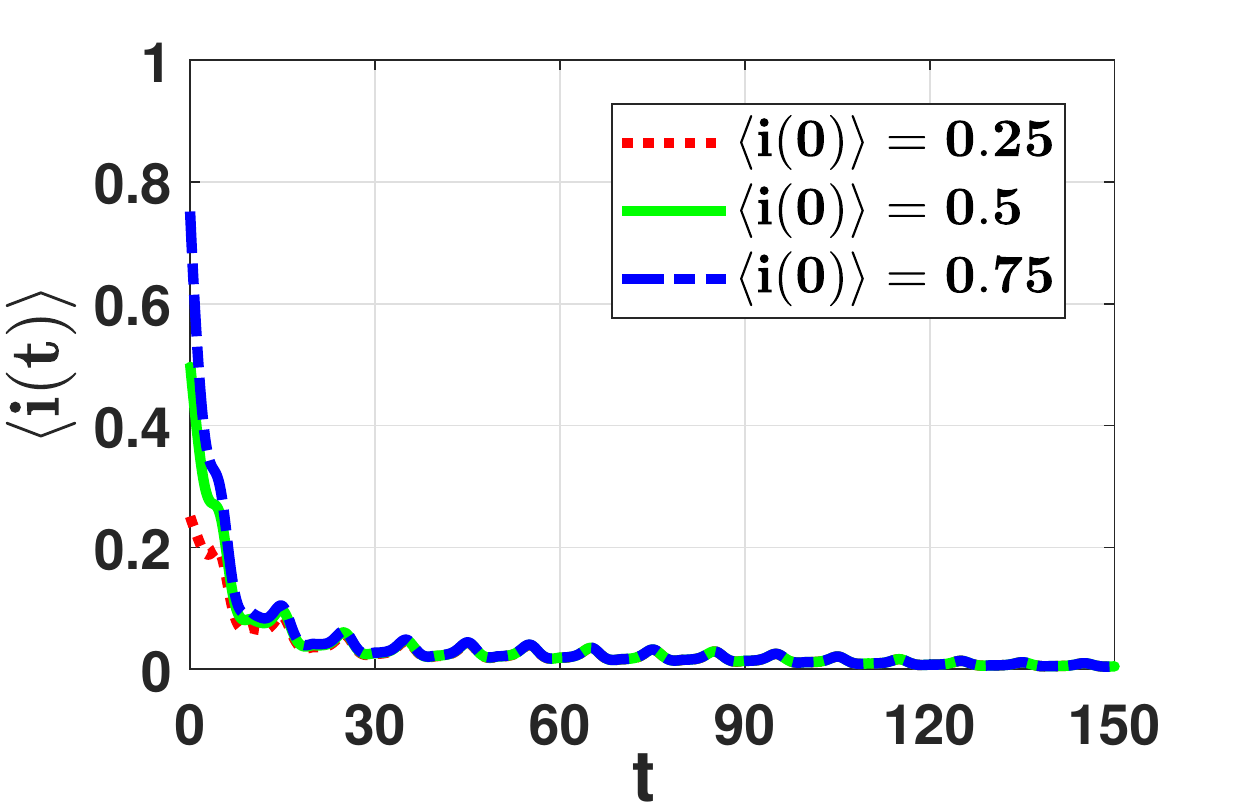}
		\caption{Parameter set (p2) }
	\end{subfigure}
	\begin{subfigure}[b]{0.23\textwidth}
		\centering
		\includegraphics[width=\textwidth]{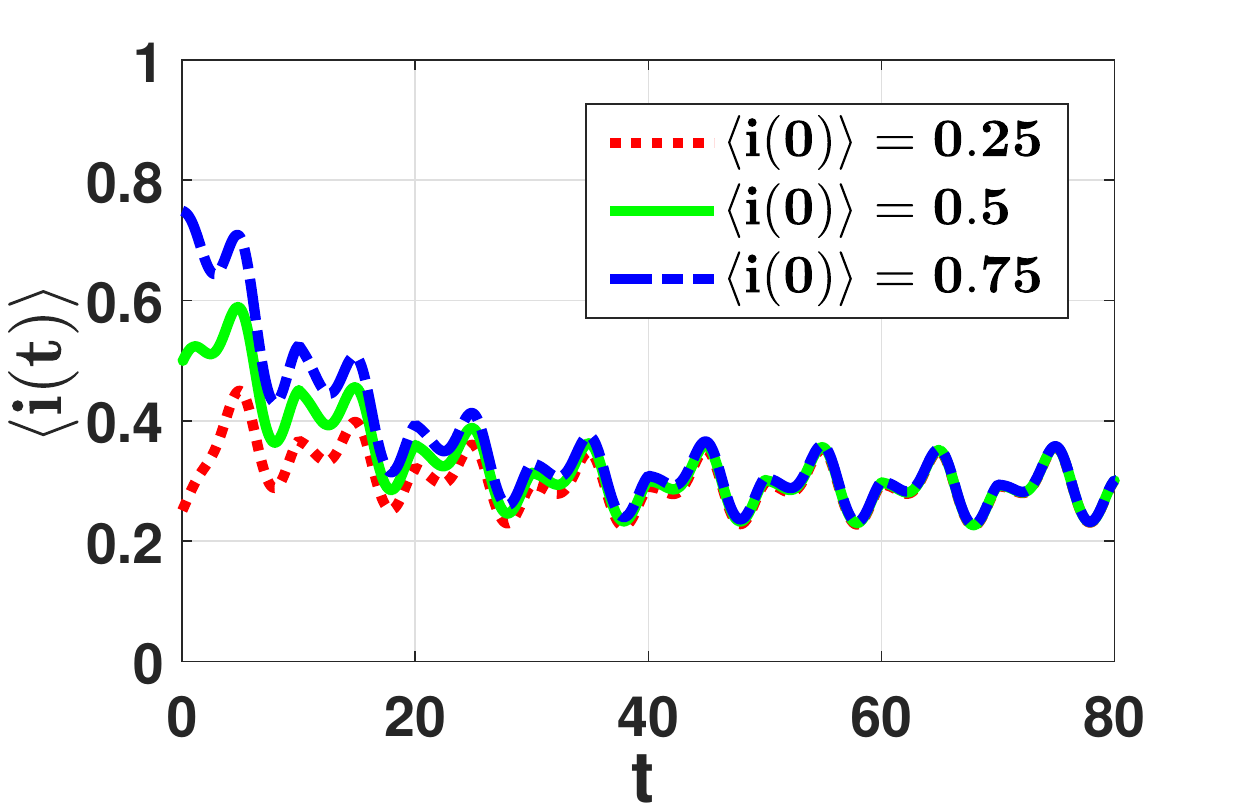}
		\caption{Parameter set (p3)}
	\end{subfigure}
	\begin{subfigure}[b]{0.23\textwidth}
		\centering
		\includegraphics[width=\textwidth]{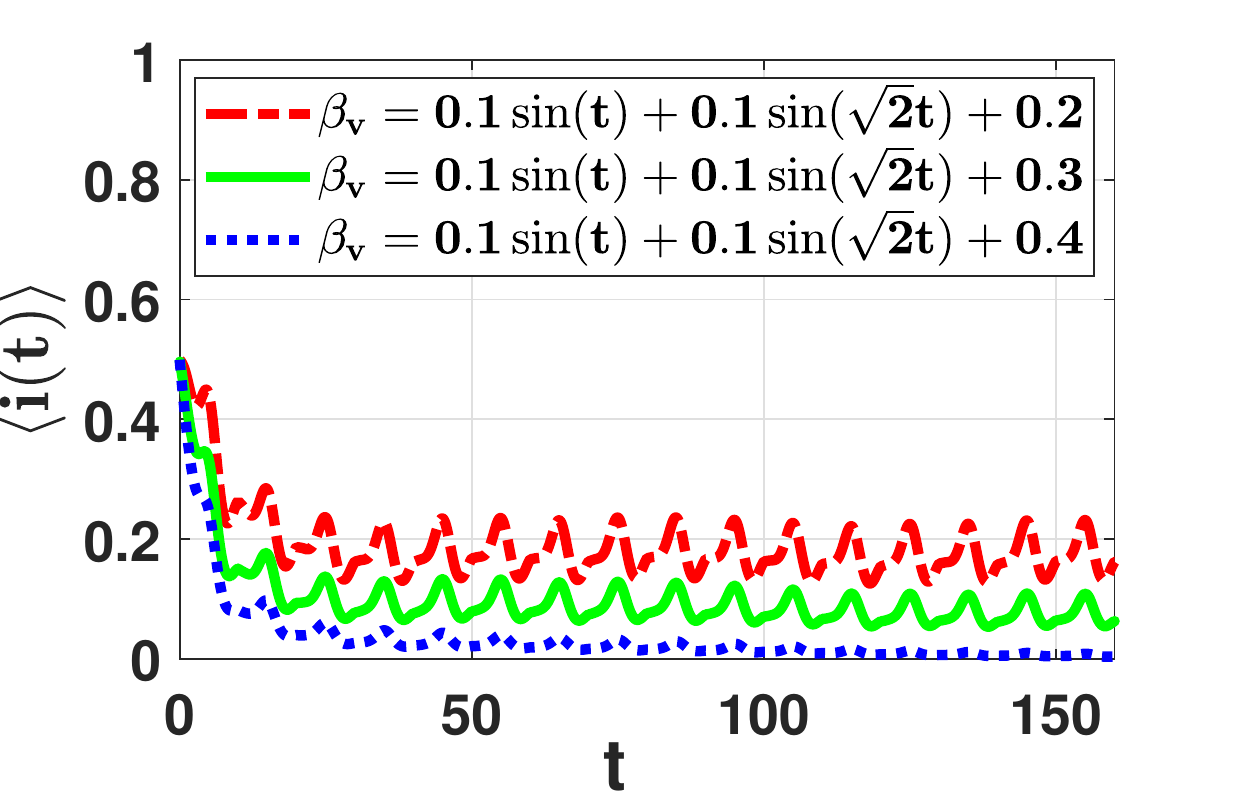}
		\caption{Parameter set (p4)}
	\end{subfigure}
	\caption{Global attractivity of the $\sum$-model (\ref{model-prod-sum}) with time-dependent parameters except that there are no pull-based attacks, namely $\alpha_v(t)=0$, $\forall v\in V$, $\forall t$.}
	\label{fig-alpha-0}
\end{figure}

Figures \ref{fig-alpha-0}(a),  \ref{fig-alpha-0}(b) and \ref{fig-alpha-0}(c) plot the dynamics of the fraction of compromised nodes over time, namely $\langle i(t)\rangle$, with parameter sets (p1), (p2), (p3) and different initial infection values. We observe that the the dynamics always converges to a unique trajectory. It reinforces the result that the dynamics converges to the equilibrium $\mathbf{0}$ when $\mu(D_if(\mathbf{0},y(t)))<0$, attracts to a positive trajectory when $\mu(D_if(\mathbf{0},y(t)))>0$, and attracts to a trajectory (possibly an equilibrium $\mathbf{0}$ or non-negative trajectory) when $\mu(D_if(\mathbf{0},y(t)))=0$. Figure \ref{fig-alpha-0}(d) plots the dynamics of $\langle i(t)\rangle$ with multiple $\beta_v(t)$'s, and shows that the globally attractive trajectory converges to $\mathbf{0}$ as $\beta_v(t)$ increases. This reinforces the intuition that a larger $\beta_v(t)$ leads to a smaller $\mu(D_if(\mathbf{0},y(t)))$ and $\langle i(t)\rangle$ as well as the convergence to the equilibrium $\mathbf{0}$.

\begin{figure}[!htbp]
	\centering
	\begin{subfigure}[b]{0.23\textwidth}
		\centering
		\includegraphics[width=\textwidth]{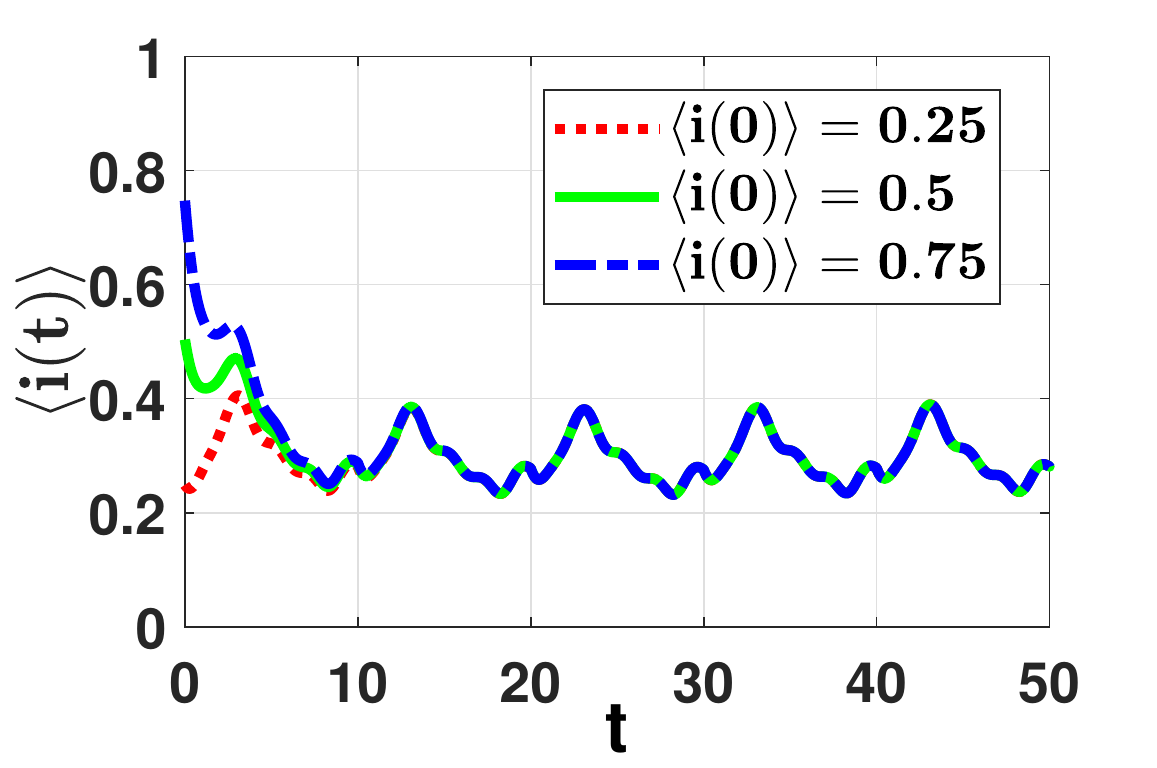}
		\caption{$|V_{\alpha>0}|/{|V|}=0.5$\label{fig-alpha-over-0-a}}
	\end{subfigure}
	\begin{subfigure}[b]{0.23\textwidth}
		\centering
		\includegraphics[width=\textwidth]{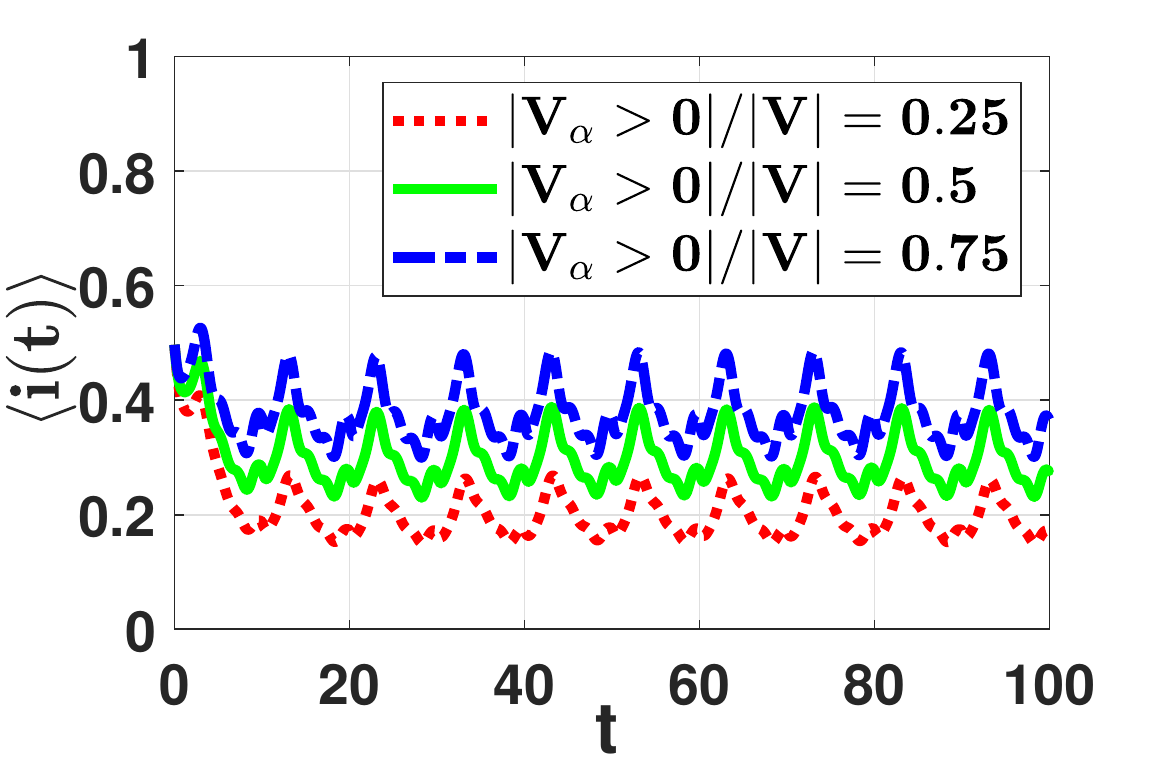}
		\caption{$\langle i(0)\rangle =0.5$\label{fig-alpha-over-0-b}}
	\end{subfigure}
	\caption{Dynamics of the $\sum$-model (\ref{model-prod-sum}) with time-dependent parameters, including different pull-based attack capabilities.}
	\label{fig-alpha-over-0}
\end{figure}

In the second experiment, we set $ \alpha_v(t) = 
0.1 \sin(3 t) + 0.1\sin(\sqrt{3} t)+0.2$ if $v\in V_{\alpha>0}$, set $\beta_v(t)$ and $\gamma_{vu}(t)$ as in the aforementioned parameter set (p2), and consider  $|V_{\alpha>0}|/{|V|}\in \{0.25,0.5,0.75\}$, namely that $25\%$, $50\%$, and $75\%$ randomly chosen nodes are subject to pull-based attacks, respectively. Figure \ref{fig-alpha-over-0}(a) plots the dynamics of $\langle i(t)\rangle$ with $|V_{\alpha>0}|/{|V|}=0.5$ and different initial values, and shows that the dynamics converges to a unique positive trajectory. Figure \ref{fig-alpha-over-0}(b) plots the dynamics of $\langle i(t)\rangle$ under different pull-based attack capabilities, and shows a positive correlation between the fraction of the compromised nodes and the pull-based attack capabilities.

\subsection{Confirming Bounds}
In order to characterize the impact of time-dependent functions and pull-based attacks on the tightness of the bounds, we consider the following four parameter sets (p5)-(p8):
\begin{itemize}
	\item (p5): $\alpha_v(t) = 0.05 \sin(3 t) + 0.05\sin(\sqrt{3} t)+0.5$, $\forall v\in V_{\alpha>0}$ where $|V_{\alpha>0}|/{|V|}=0.5$; $\beta_v(t)= 0.05\sin(t) + 0.05 \sin(\sqrt{2} t)+0.5$, $\forall v\in V$; $\gamma_{uv}(t)=0.1\sin(\pi t/5)+0.5$, $\forall (u,v)\in E(t)$;
	
	\item (p6): $\alpha_v = 0$, $\beta_v(t)= 0.1\sin(t) + 0.1 \sin(\sqrt{2} t)+0.4$, $\forall v\in V$; $\gamma_{uv}(t)=0.1\sin(\pi t/5)+0.5$, $\forall (u,v)\in E(t)$;
	
	\item (p7): $\{\alpha_v(t)\}_{t\ge 0}\sim U([0.1,0.3])$, $\forall v\in V_{\alpha>0}$ where $|V_{\alpha>0}|/{|V|}=0.5$; $\{\beta_v(t)\}_{t\ge 0}\sim U([0.4,0.7])$, $\forall v\in V$; $\gamma_{uv}(t)=0.1$, $\forall (u,v)\in E(t)$;
	
	\item (p8): $\alpha_v = 0$, $\{\beta_v(t)\}_{t\ge 0}\sim U([0.4,0.7])$, $\forall v\in V$; $\gamma_{uv}(t)=0.1$, $\forall (u,v)\in E(t)$.	
\end{itemize}

Note that in these settings, it holds that $\{\eta(t)\}_{t\ge 0}\sim U([a,b])$ if $\eta(t)$ satisfies (i) $\eta(t) = \eta_{k}$ when $t\in [k,k+1), k\in \N$ and (ii) $\eta_{k},k\in\N$ follow the uniformly distribution in $[a,b]$ independently. Moreover, parameter sets (p7) and (p8) satisfy the ergodic property.

\begin{figure}[!htbp] 
	\centering
	\begin{subfigure}[b]{0.23\textwidth}
		\centering
		\includegraphics[width=\textwidth]{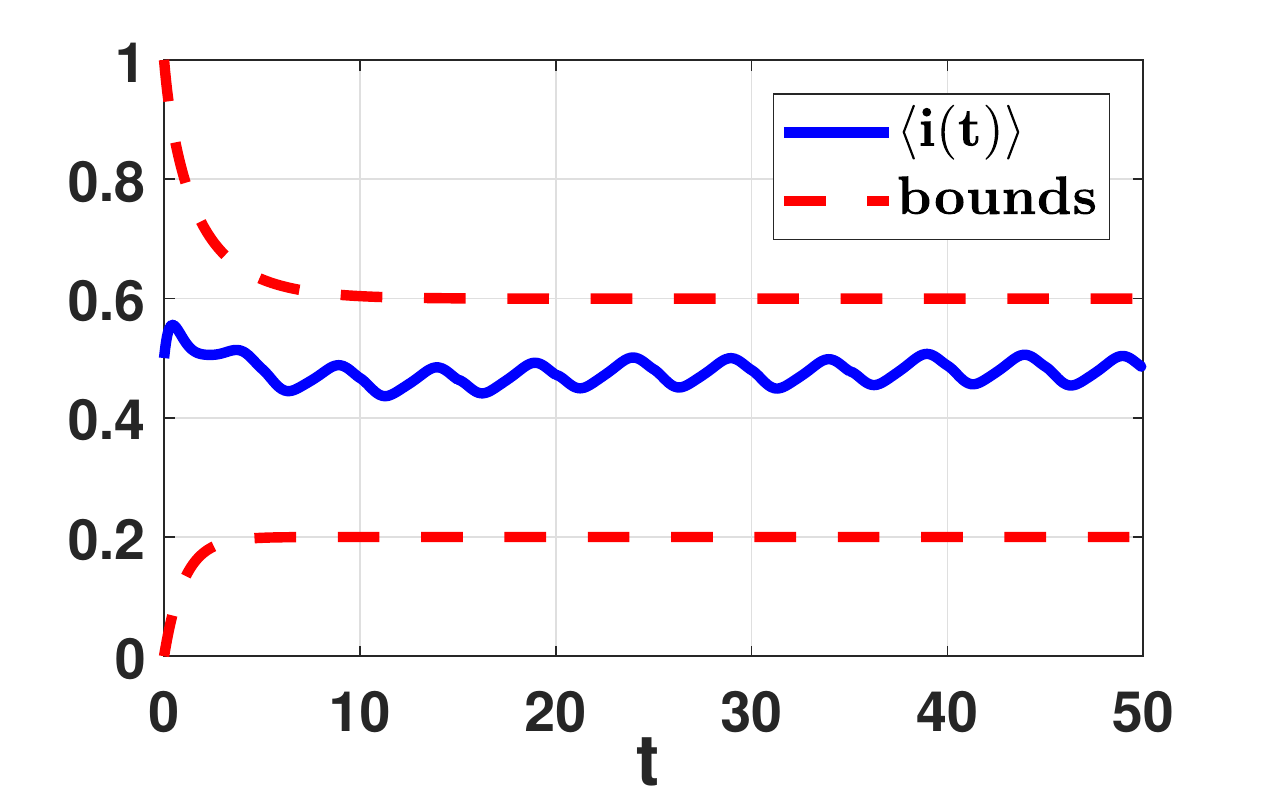}
		\caption{Parameter set (p5)}
	\end{subfigure}
	\begin{subfigure}[b]{0.23\textwidth}
		\centering
		\includegraphics[width=\textwidth]{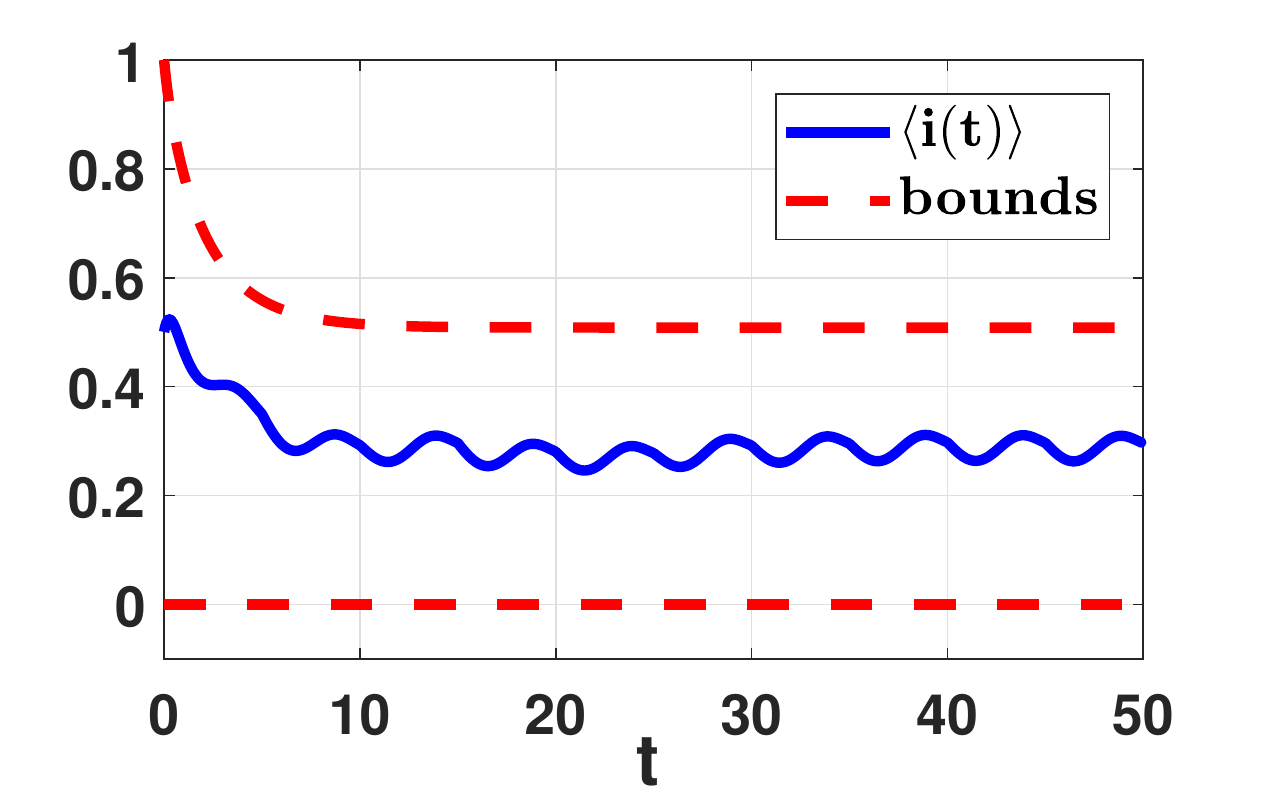}
		\caption{Parameter set (p6)}
	\end{subfigure}
	\begin{subfigure}[b]{0.23\textwidth}
		\centering
		\includegraphics[width=\textwidth]{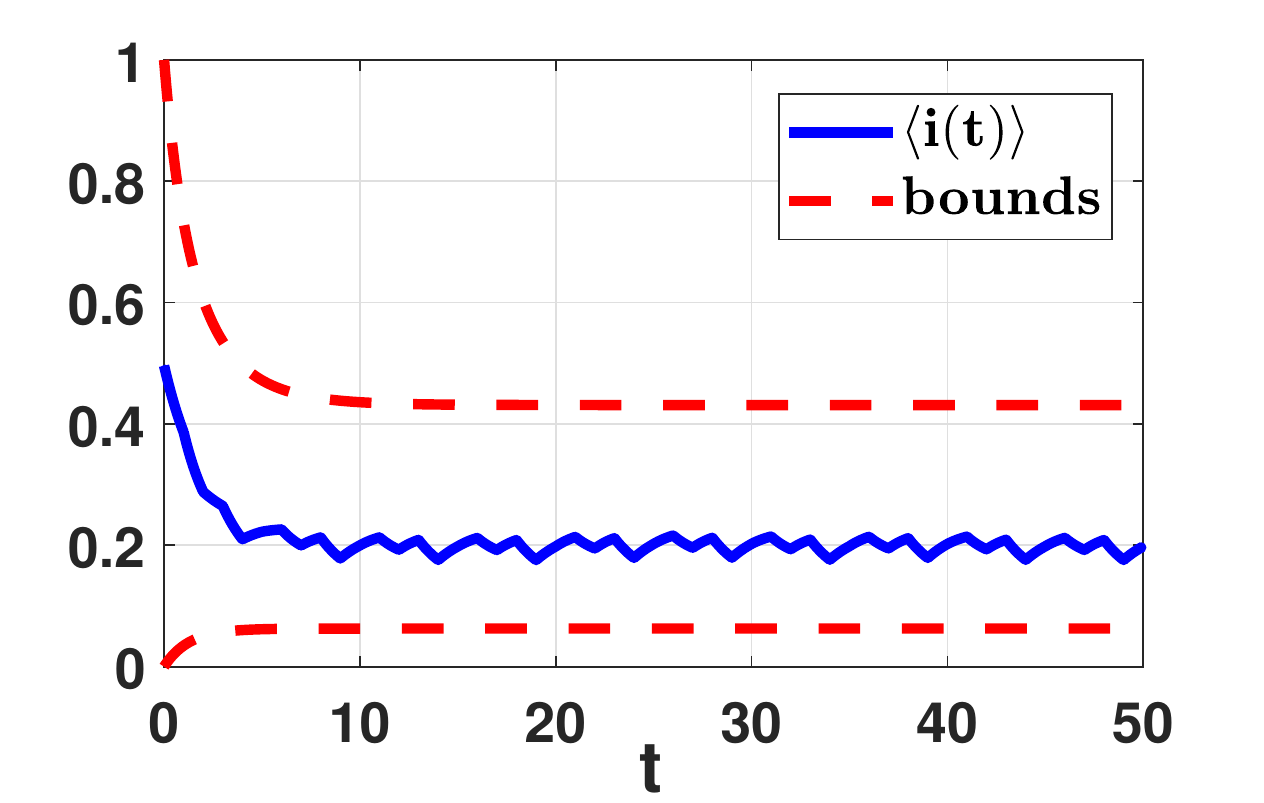}
		\caption{Parameter set (p7)}
	\end{subfigure}
	\begin{subfigure}[b]{0.23\textwidth}
		\centering
		\includegraphics[width=\textwidth]{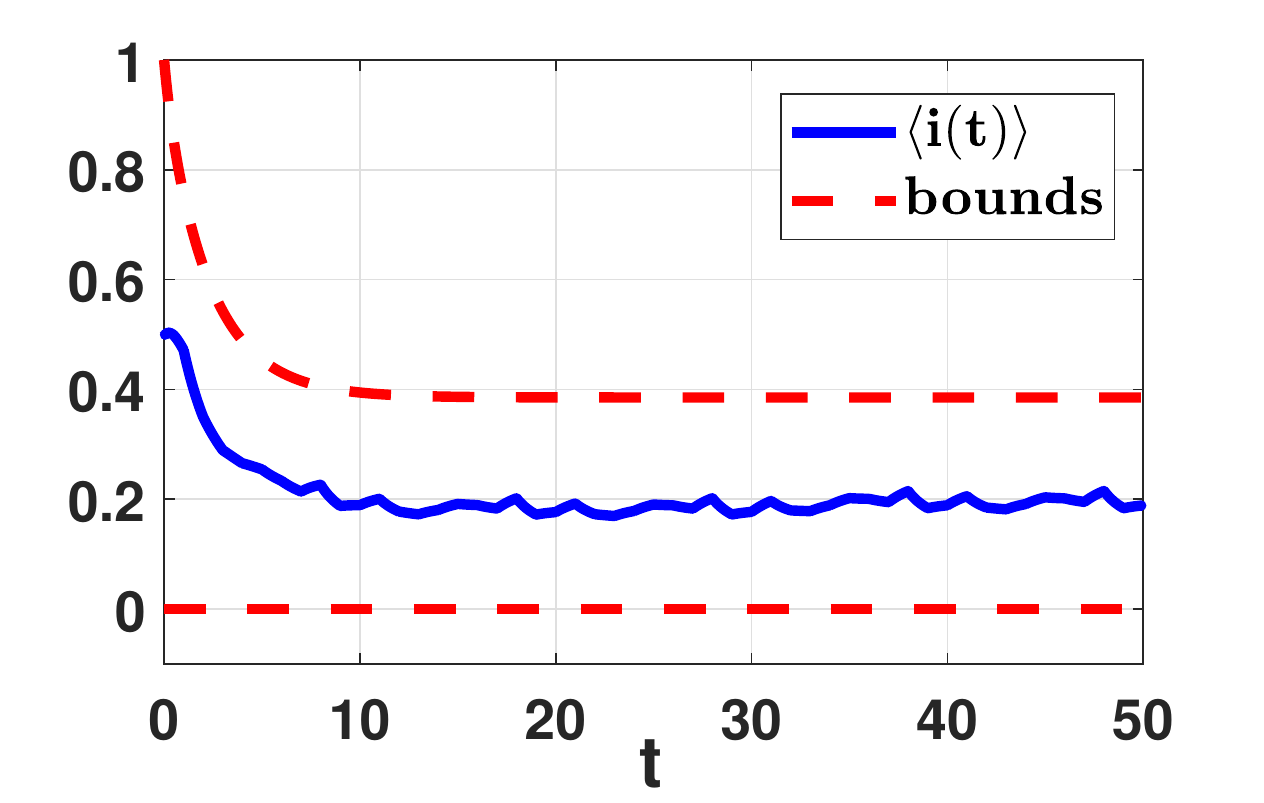}
		\caption{Parameter set (p8)}
	\end{subfigure}
	\caption{Dynamics of the $\sum$-model 
		(\ref{model-prod-sum}) with time-dependent parameters and their bounds given by Theorem \ref{thm-bounds}.}
	\label{fig-bounds}
\end{figure}

Figure \ref{fig-bounds} plots the dynamics of $\langle i(t)\rangle$ with parameter sets (p5)-(p8) and the corresponding bounds given by Theorem \ref{thm-bounds}. The lower bounds in Figure \ref{fig-bounds} (b) and (d) are $0$ because $i_{\min}=\mathbf{0}$ and $\alpha_v(t)=0$ in parameter sets (p6) and (p8). We observe that the bounds in Figure \ref{fig-bounds} can be very loose, perhaps because $i_{\max}=\mathbf{1}$ and $i_{\min}=\mathbf{0}$ in this example, which reflects the lack of information on the globally attractive trajectory.

\subsection{Are the Sufficient Conditions Necessary?} \label{subsec:necessity}

It is known \cite{XuHotSoS15} that some cybersecurity dynamics can exhibit bifurcation and chaos. Since we leverage subhomogeneity (Property \ref{pro-subhomo}) and ergodicity (Property \ref{property-ergodic}) to obtain the global attractivity result for the unified dynamics with time-dependent parameters, it makes us wonder how far these sufficient conditions are from being necessary. In what follows we use numerical examples to show that violating these properties can cause violation of global attractivity, hinting subhomogeneity and ergodicity may be necessary; the rigorous treatment of this is a difficult task and left for future research.

For constructing examples, it suffices to consider a special kind of time-independent attack-defense structures $G$ in Erd\"os-R\'enyi (ER) random graph. Specifically, we consider an ER structure with $n=1,000$ nodes and edge probability $p=0.1$. The ER graph is then interpreted as a directed graph. We consider the $\sum$-model as an example.

First, we empirically show that subhomogeneity (required by Property \ref{pro-subhomo}) may be necessary for global attractivity. Let us consider the following functions for the $\sum$-model:
{\small
	\begin{eqnarray*}
		h_v(i,\beta_v) &=& \beta_v(t),\\ 
		g_v(i,\alpha_v,\Gamma) &=& \alpha_v(t) + \left(\frac{\sum_{u\in N_v(t)}\gamma_{vu}(t) a_{vu}(t) i_u(t)}{\max\{|N_v(t)|,1\}}\right)^2.
	\end{eqnarray*}
}
Note that the preceding $g_v$ is {\em not} subhomogeneous. We consider the following two combinations: 
\begin{itemize}
	\item (p9): $\alpha_v=0$, $\forall v$;
	$\beta_v(t)= 0.05\sin(t) + 0.05 \sin(\sqrt{2} t)+0.1$, $\forall v\in V$; $\gamma_{uv}(t)=0.1\sin(\pi t/5)+0.7$, $\forall (u,v)\in E(t)$;
	\item (p10): $\alpha_v(t)=0.1 \sin(3t) + 0.1\sin(\sqrt{3} t)+0.1$, $\forall v\in V_{\alpha> 0}$ where $|V_{\alpha> 0}|/|V| = 0.2$; the other parameters are the same as in (p9). 
\end{itemize}

\begin{figure}[!htbp]
	\centering
	\begin{subfigure}[b]{0.23\textwidth}
		\centering
		\includegraphics[width=\textwidth]{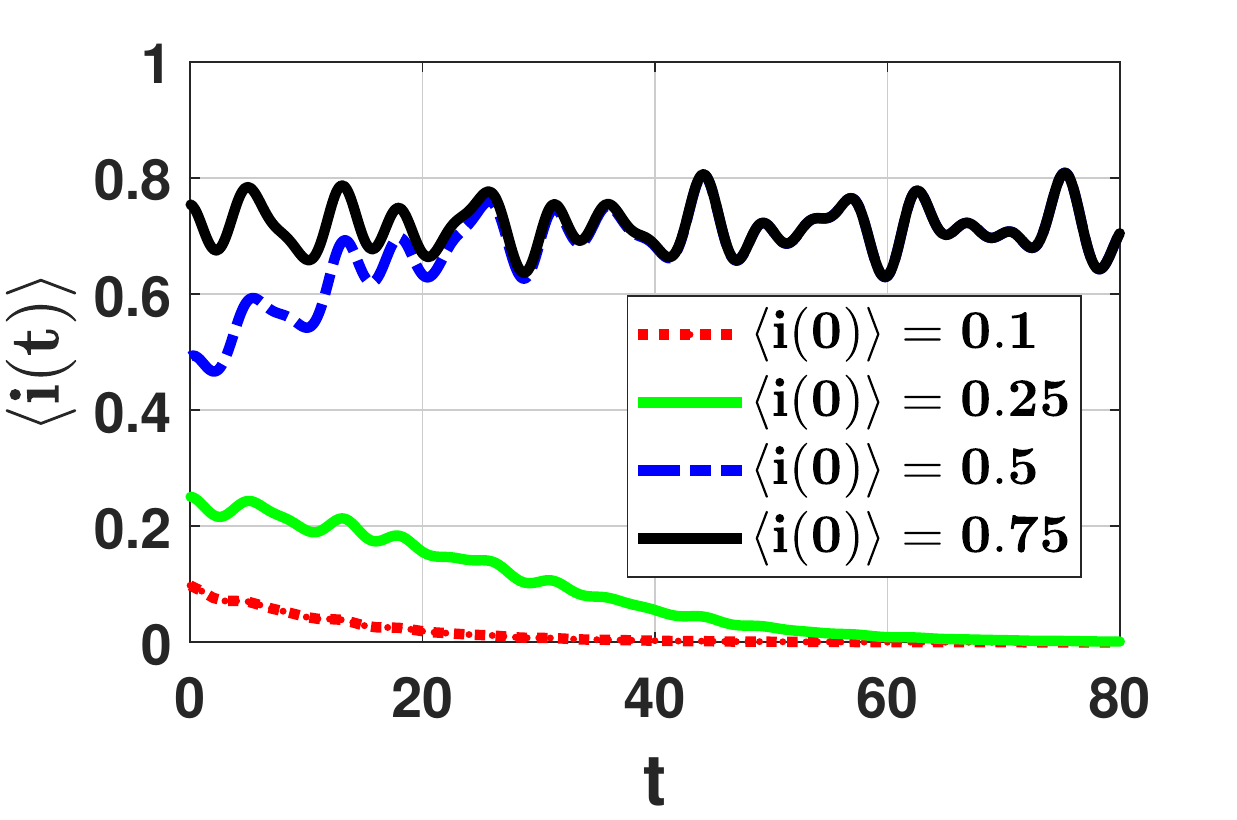}
		\caption{{\footnotesize (p9) with $|V_{\alpha>0}|/{|V|}=0$}}
	\end{subfigure}
	\begin{subfigure}[b]{0.23\textwidth}
		\centering
		\includegraphics[width=\textwidth]{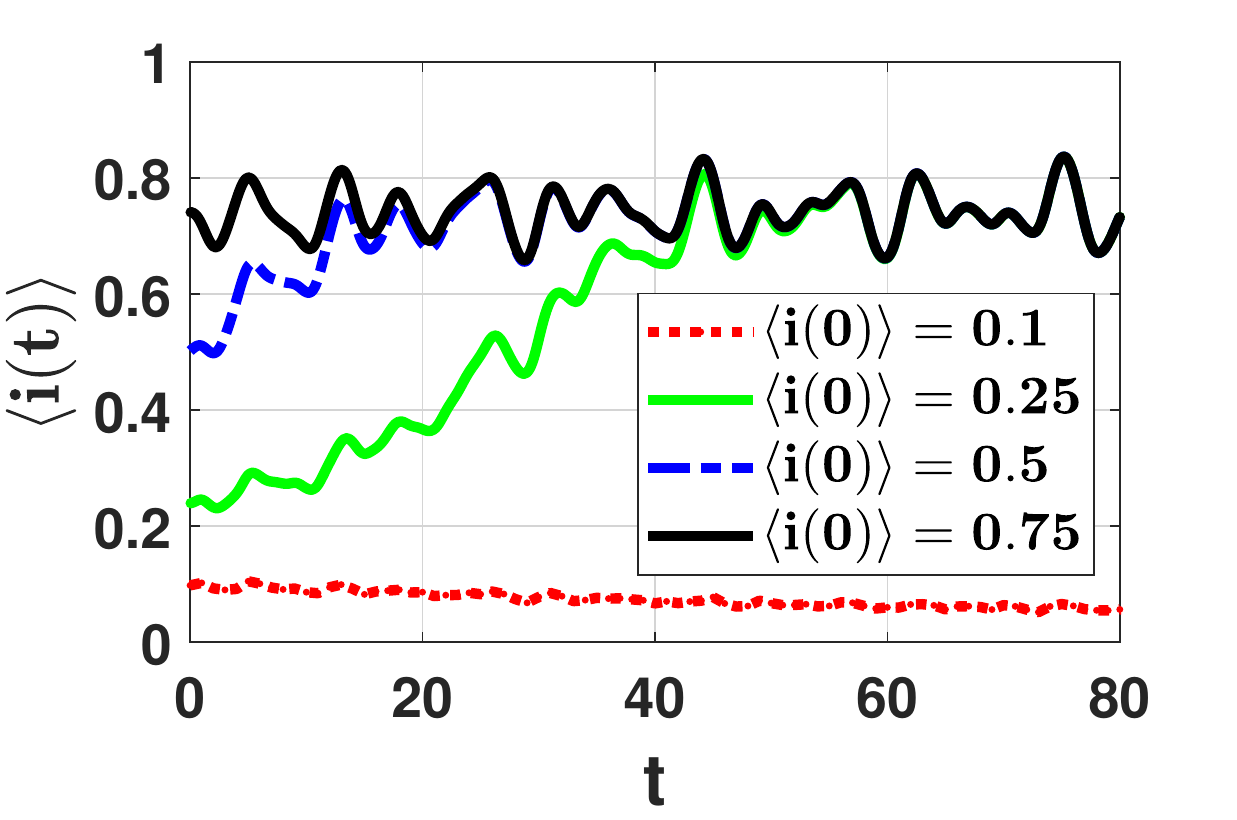}
		\caption{{\footnotesize (p10) with $|V_{\alpha>0}|/{|V|}=0.2$}}
	\end{subfigure}
	\caption{Examples showing that the $\sum$-model \eqref{model-prod-sum} is not globally attractive when $g_v$ is not subhomogeneous.}
	\label{fig-fail-converge}
\end{figure}

Figure \ref{fig-fail-converge}(a) plots the dynamics of $\langle i(t)\rangle$ with parameter set (p9) but different initial values $\langle i(0)\rangle$. We observe that the dynamics is not globally attractive because different initial values lead to different trajectories. Figure \ref{fig-fail-converge}(b) plots the dynamics of $\langle i(t)\rangle$ with parameter set (p10) but different initial values $\langle i(0)\rangle$. We observe that the dynamics is not globally attractive because different initial values lead to different trajectories. These experiments hint that subhomogeneity may be necessary for global attractivity.

\begin{SCfigure}
	\caption{Example with parameter set (p11) showing that the $\sum$-model	\eqref{model-prod-sum} is not globally attractive when $\{\beta_v(t)\}_{t\geq 0}$ is not ergodic.}
	\label{fig-non-ergodic}
	\includegraphics[width=0.26\textwidth]{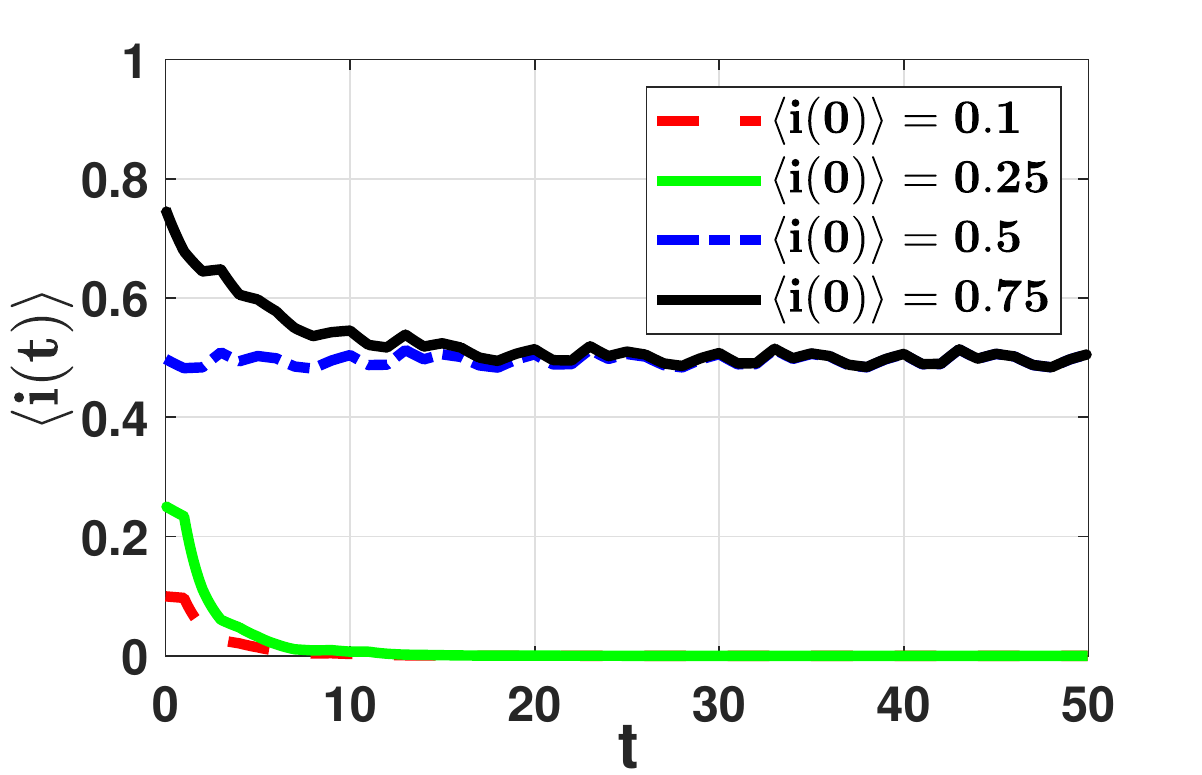}
	\vspace{-0.1in}
\end{SCfigure}

Second, we empirically show that ergodicity may be necessary for global attractivity. Let us consider the $\sum$-model \eqref{model-prod-sum} without pull-based attacks, namely
\begin{itemize}
	\item (p11): $\alpha_v(t)=0$, $\forall v\in V$;  $\gamma_{uv}(t)=0.3$, $\forall (u,v)\in E(t)$ and $t\ge 0$;
	$$
	\{\beta_v(t)\}_{t\ge 0}\sim \left\{\begin{array}{cc}
	U([0.1,0.2])  & {\rm~with~probability~1/2}, \\
	U([0.1,1])  & {\rm~with~probability~1/2}.
	\end{array}
	\right.
	$$
\end{itemize}
Note that $\{\beta_v(t)\}_{t\ge 0}$ is not ergodic in this example.
Figure \ref{fig-non-ergodic} plots the dynamics of $\langle i(t)\rangle$ with parameter set (p11) but  different initial values $\langle i(0)\rangle$ as well as different realizations of the non-ergodic $\{\beta_v(t)\}_{t\geq 0}$. We observe  that the dynamics is not globally attractive because different initial values lead to different trajectories, hinting that ergodicity may be necessary.

\section{Conclusion}\label{sec-conclusion}

We have proved that preventive and reactive cyber defense dynamics with ergodic time-dependent parameters is globally attractive and the dynamics is further (almost) periodic when the time-dependent parameters are (almost) periodic. 
These theoretical results supersede the state-of-the-art understanding of at least two models extensively investigated in the literature, and shed a light on the boundary between ``when the dynamics is analytically treatable'' and ``when the dynamics is not analytically treatable''.
There are important,  but challenging, open problems for future research. First, we numerically showed that 
ergodicity may be necessary for global attractivity.
It is therefore important to rigorously pin down the necessary conditions under which the dynamics is globally attractive, namely the precise boundary between ``when the dynamics is analytically treatable'' and ``when the dynamics is not analytically treatable''.
Second, we did not characterize the convergence speed, which is another challenging problem because the globally attractive trajectory is time-dependent, rendering the eigenvalue analysis of Jacobian matrix not applicable here. 

\bibliographystyle{IEEEtran}
\bibliography{complex-network}

\newpage
~
\newpage

\appendices
\section{Proof of Lemma \ref{lemma-linear-sys}}
\label{appendix:lemma-linear-sys}

\begin{proof}
	For proving (i), we observe that the boundedness of $B(t)$ means that there exists $M_1>0$ such that $\sum_{k=1}^n b_{jk}(t)\ge-M_1$ holds for $\forall j$, $\forall t$. Let $z_{i_0(t)}(t) = \min_{i} z_i(t)$. It follows
	\begin{align}\nonumber
	\frac{dz_{i_0(t)}}{dt} =& b_{i_0(t) i_0(t)}(t) z_{i_0(t)}(t) + \sum_{j\neq i_0(t)} b_{i_0(t) j}(t) z_j(t)\\\nonumber
	\ge &b_{i_0(t) i_0(t)}(t) z_{i_0(t)}(t) + \sum_{j\neq i_0(t)} b_{i_0(t) j}(t) z_{i_0(t)}(t) \\\label{element-nonegative}
	\ge &-M_1 z_{i_0(t)}(t);
	\end{align}
	this implies $z_{i_0(t)}(t)\ge 0$ when $z(0)\ge\mathbf{0}$. This means that the solution matrix $U(t,s)$ is nonnegative for any $t\ge s\ge 0$. 
	It follows from the ergodicity of $\{B(t)\}_{t\in\R}$ that the convergence 
	\begin{align}\label{eq:mean-B}
	\lim_{t\to\infty} \frac{1}{t}\int_{a}^{a+t} B(s)ds = M(B)
	\end{align}
	holds uniformly for any $a\in\R$. 
	
	For proving (ii), we observe that  if $\mathcal G(M(B))$ is strongly connected, then there exists $\delta> 0$ such that 
	$\delta$-graph of $M(B)$ is strongly connected. It then follows from Eq. (\ref{eq:mean-B}) that there exists $T_1>0$, such that for any interval $[kT_1,(k+1)T_1)$ for any  $k\in\N$, the $\frac{\delta}{2}$-graph of  $\int_{kT_1}^{(k+1)T_1} B(s)ds$ is strongly connected. According to the proof of Lemma 2(2) in \cite{han2015achieving}, there exists $\eta_1>0$ such that the $\eta_1$-matrix of $U((k+1)T_1,kT_1)$ is irreducible. On the other hand, Inequality \eqref{element-nonegative} implies that 
	\begin{align}\label{solution-matrix-diagonal}
	U_{jj}(t,s)\ge e^{-M_1(t-s)}, \forall j.
	\end{align}
	Following the proof of Lemma 1 in \cite{han2013cluster}, we know that the product of $(n-1)$-many $n$-dimensional nonnegative matrices, which are irreducible and have positive diagonal elements, gives a positive matrix. 
	This means that there exists $\eta_2>0$ such that 
	\begin{align} \label{product-solution-matrices}
	U(t_{k+n},t_{k+1}) = U(t_{k+n},t_{k+n-1})\cdots U(t_{k+2},t_{k+1})>\eta_2
	\end{align}
	for $\forall k\in\N$. Let $T = nT_1$ and $t_k = kT$, for any $k\in\N$. Then by Inequalities (\ref{solution-matrix-diagonal}) and (\ref{product-solution-matrices}), the following holds for any $\Delta>T$ and any $s\in [t_N,t_{N+1})$, $\forall N\in \N$:
	\begin{align*}
	U(s+\Delta,s) = &U(t_{N+1},s)U(t_{N+n},t_{N+1})U(s+\Delta,t_{N+n})\\
	\ge &\eta_2 e^{-M_1(t_{N+1}-s)}e^{-M_1(s+\Delta-t_{N+n})}\ge \eta_2 e^{-M_1\Delta}.
	\end{align*}
	By letting $\epsilon = \eta_2 e^{-M_1\Delta}$, we complete the proof.
\end{proof}

\section{Proof of Theorem \ref{thm-alpha-0-stability}}
\label{appendix:thm-alpha-0-stability}

\begin{proof}
	It follows from Property \ref{pro-subhomo} and Theorem \ref{thm-pre-subhomogeneous} that $f(i,y)$ is {\em strongly subhomogeneous}, meaning that for any $\alpha>0$ and $i\in [0,1]^n$, we have
	\begin{align*}
	f(i,y) \le \frac{f(\alpha i,y)}{\alpha} &= \frac{f(\mathbf{0},y)+\alpha D_i f(\mathbf{0},y)i + o(\|\alpha i\|)}{\alpha}\\
	& =  D_i f(\mathbf{0},y)i + \frac{o(\|\alpha i\|)}{\alpha}.
	\end{align*}
	When $\alpha\to 0$, we have $f(i,y)\le  D_i f(\mathbf{0},y)i$. This means that for $i \in \R^n_{\ge 0}$, system $dz(t)/dt= D_i(\mathbf{0},y(t)) z(t)$ is a comparison model to model \eqref{eq:unified-generic-model}. If $i(0) = z(0)\in [0,1]^n$, $i_v(t)\le z_v(t)$ holds for $\forall t>0, \forall v$ .  
	
	When $\mu(D_i f(\mathbf{0},y))<0$, it follows that $\forall v$, $\lim_{t\to\infty}z_v(t)= 0$ holds for any initial value $z(0)\in  [0,1]^n$. The global convergence to the equilibrium $\mathbf{0}$ follows from the fact $i_v(t)\le z_v(t)$.
\end{proof}

\section{Proof of Lemma \ref{lemma-persistent}}
\label{appendix:lemma-persistent}

\begin{proof}
	Consider the linear variational equation of model (\ref{eq:unified-generic-model}) at the equilibrium $\mathbf{0}$,
	\begin{align}\label{var}
	\frac{dz}{dt}=D_i f(\mathbf{0},y(t))z,
	\end{align}
	where $D_i\psi(t,\mathbf{0},y)$ is the fundamental solution matrix.
	Since $\{y(t)\}_{t\ge 0}$ is ergodic and the Jacobian matrix $D_i f(\mathbf{0},y))$ is bounded, Oseledets multiplicative ergodic theorem of random dynamical systems (Theorems 3.4.1 and 3.4.11 in \cite{arnold2013random}) says that there exists an invariant set $\tilde{Y}\in\F$ of full measure on which
	there is an Oseledets splitting 
	$\R^n = E_{1}(y)\oplus \dots\oplus E_r(y)$
	associated with $\rho_1> \dots> \rho_r$ such that for $k=1,\cdots,r$,
	$$
	D_i\psi(t,\mathbf{0},y) E_{k}(y) = E_{k}(\theta(t,y))
	$$
	and 
	\begin{align}\label{eq-le}
	\lim_{t\to\infty}\frac{1}{t}\log\|D_i\psi(t,\mathbf{0},y) \xi\| = \rho_k,~{\rm for}~\xi\in E_{k}(y)\backslash\{\mathbf{0}\},
	\end{align}
	where convergence is uniform in $E_k(y)\cap S_1^{n}$ with $S_{1}^{n}=\{x\in\R^{n}: \|x\| = 1\}$.  On the other hand, it follows from Properties \ref{pro-continuous}-\ref{pro-mono} and Lemma \ref{lemma-linear-sys} that $D_i\psi(t,\mathbf{0},y) $ is nonnegative for all $t\in\R$, $y\in Y$ and $\R_{\ge 0}^{n}$ is invariant with respect to model \eqref{var}. Hence, $E_{1}(y)\subseteq \R_{\ge 0}^{n}$.
	
	Consider the dual system of model \eqref{var} with $s<t$ as follows
	\begin{align}
	\frac{dx(s)}{ds} = -x(s)D_i f(\mathbf{0},y(s)),\label{var1}
	\end{align}
	to which the fundamental solution matrix is $U_1(s,t)$. Then, we have $U_1(s,t)=U(t,s)$. It can be seen that model \eqref{var1} has Lyapunov exponents $-\rho_{j}, j=1,\dots,r$. Denote the associated Oseledets splitting subspaces by $F_{j}(y)$,  $j=1,\dots,r$. Since $\R_{\le 0}^{n}$ is invariant for model (\ref{var1}), $F_{1}(y)\in \R_{\le 0}^{n}$ for all $x$ and $y$.
	
	Consider $
	\varphi_{1}^{\top}(t)U(t,s)\phi_{j}(s)
	$
	with $\phi_{j}(s)\in E_{j}(\theta(s,y) )\cap S_1^{n}$ and $\varphi_{1}(t)^{\top}\in F_{1}(\theta(t,y))\cap S_1^{n}$ for any $j\ne 1$. As $t-s\to+\infty$, it can be respectively approximated as follows: 
	\begin{align*}
	&\varphi_{1}^{\top}(t)U(t,s)\phi_{j}(s)\sim\exp(\rho_{j}(t-s))\varphi_{1}^{\top}(t)\phi_{j}(t),\\
	&\varphi_{1}^{\top}(t)U(t,s)\phi_{j}(s)\sim\exp(-\rho_{1}(s-t))\varphi_{1}^{\top}(s)\phi_{j}(s),
	\end{align*}
	which implies
	\begin{align*}
	\varphi_{1}^{\top}(t)\phi_{j}(t)\sim\exp((\rho_{1}-\rho_{j})(t-s))\varphi_{1}^{\top}(s)\phi_{j}(s).
	\end{align*}
	Since $\varphi_1(t)$ and $\phi_j(t)$ are bounded, we have $\varphi_{1}^{\top}(t)\phi_{j}(t)=0$ for all $t$ and $j\ne 1$. On the other hand, from Property \ref{pro-mono} and $\mathcal G(M(\Gamma))$ being strongly connected, we know $\mathcal G(M(D_i f(\mathbf{0},y)))$ is strongly connected. Then, it follows from Lemma \ref{lemma-linear-sys} that $U(t,s)$ is non-negative and there exist $\Delta>0$ and $\epsilon_1>0$ such that every element of $U(t+\Delta,t)$ is greater than $\epsilon_1$. 
	This means that $\varphi_1(t)\in\R_{\ll 0}^n\cap S_1^n$ when $t\ge \Delta$. Then, from the compactness of $[0,1]^n$ and $Y$, we have that the angles $\angle(E_j(y),\R_{>0}^n) >\eta$ for $j\neq 1$ and some $\eta>0$.
	Then, we have that there exists $M>0$ such that for any $\xi\in\R_{>0}^n$,
	$$
	\|D_i \psi(t,\mathbf{0},y)\xi\| \ge  M\|\xi\| \exp(r t)
	$$
	holds for some $r>0$. Since Property $\mathbf{1}$ says that $f(i,y)$ has continuous first and second derivatives with respect to $i$ and model \eqref{var} is the first order linear approximation to model \eqref{eq:unified-generic-model} around the equilibrium $\mathbf{0}$, there exists $\epsilon>0$ and $T'>0$ such that for any $i_0\in\R_{> 0}^n$, we have 
	\begin{align}\label{extract}
	\|\psi(t,i_0,y)\|> \epsilon
	\end{align}
	when $t> T'$. For an index set $C\subseteq \{1,\cdots,n\}$, let 
	$$Y^C_{\epsilon} = \{x\in\R_{\ge 0}^n: x_j=0,~\forall~j\in C~{\rm and}~x_{j'}>\epsilon,~\forall~j'\notin C\}.$$ 
	Suppose $\psi(t,i_0,y)$ goes near $Y^C_{\epsilon}$ and let $z(t) = [i(t)]_{C}\in\R^{|C|}$, it follows from the mean value theorem that
	\begin{align}\label{eq:linear-var}
	\frac{dz(t)}{dt}= A(t)z(t)+I(t),
	\end{align}
	where $A(t) = [D_i f(\hat{i},y)]_{C,C}$ and $I(t)= [f(\tilde{i},y)]_C$ with $\tilde{i}\in Y_{\epsilon}^C$ and $\hat{i} = c \tilde{i}+(1-c) i$, $c\in(0,1)$. From the definition of $Y^C_{\epsilon}$ and the strong connectivity of $\mathcal G(M(\Gamma))$, we know there exists $T_1>0$ such that 
	$\max_j \int_{t}^{t+T_1} I_j(t)dt>\epsilon'$ for some $\epsilon'>0$ and any $t>0$. Let $U_2(t,s)$ denote the fundamental solution matrix of system $du(t)/dt = A(t) u(t)$. Lemma \ref{lemma-linear-sys} says that there exists $T_2>0$ such that for each $\Delta>T_2$, one can find $\epsilon''(\Delta)>0$ such that $U_2(t+\Delta,t)\ge \epsilon''(\Delta)$. This means that the solution to model (\ref{eq:linear-var}) with non-negative initial value, denoted by $z(t)$, satisfies the following: For $t>T_1+T_2+1$,
	\begin{small}
		\begin{align*}
		z(t)\ge &\int_{0}^t U_2(t,s)I(s)ds \ge \int_{t-T_1-T_2}^{t-T_2} \epsilon''(t-s) I(s)ds\\
		\ge & \min_{\Delta\in[T_2,T_1+T_2]}\epsilon''(\Delta) \int_{t-T_1-T_2}^{t-T_2} I(s) ds \ge \epsilon'\min_{\Delta\in[T_2,T_1+T_2]}\epsilon''(\Delta).
		\end{align*}
	\end{small}
	This implies that for each $C\subset\{\oneton\}$ and a sufficiently small $\epsilon>0$, there exists $\upsilon_{C,\epsilon}>0$ such that for each $i_{C,0}\in Y^{C}_{\epsilon}$, there exists $T>0$ so that $\|[\psi(t,i_{C,0},y)]_{C}\|>\upsilon_{C,\epsilon}$ for each $t>T$.
	
	Now we can complete the proof of persistence by induction. First, inequality \eqref{extract} implies that the trajectory $\psi(t,i_0,y)$ essentially goes out of the ball $B^{n}_{\epsilon}=\{x\in\R^{n}_{\ge 0}:~\|x\|\le \epsilon\}$. So, there exists at least one index, say $j(t)$, such that $[\psi(t,i_0,y)]_{j(t)}>\epsilon_{1}=\epsilon/n$ for any sufficiently large $t$. If $\psi(t,i_0,y)$ goes near $Y^{C_{{j}}}_{\epsilon_{1}}$ where $C_{{j}}=\{\oneton\}-\{j(t)\}$, the preceding analysis indicates that $[\psi(t,i_0,y)]_{C_{{j}}}$ essentially goes out of the ball $B^{C_{j}}_{\epsilon_{1}'}=\{x:~\|x_{C_{j}}\|\le\epsilon_{1}'\}$.
	
	Then, by induction, we can prove that $\psi(t,i_0,y)$ essentially goes out of the union of the following balls
	\begin{align*}
	\bigcup_{C\subset\{\oneton\}}B^{C}_{\epsilon_{C}},
	\end{align*}
	meaning that $\psi(t,i_0,y)$ is persistent for any $i_0\in[0,1]^n\bo$. 
	Now it can be concluded that $\psi(t,i_0,y)$ is persistent for any $i_0\in[0,1]^n\bo$ and almost all $y\in Y$.
\end{proof}

\section{Proof of Lemma \ref{lemma-alpha-0-ergodic}}
\label{appendix:lemma-alpha-0-ergodic}

\begin{proof}
	First, we prove part (i). Let $Y$ denote the parameter space. Note that when $y\in Y$ is ergodic and $Y$ is compact, the conditions on $Y$ in Theorem \ref{thm-pre-global-attractivity} are satisfied. Therefore, 
	the global attractivity result can be derived by employing Theorem \ref{thm-pre-global-attractivity} when the dynamics is uniformly persistent and satisfies Conditions (I) and (II) in Theorem \ref{thm-pre-global-attractivity}. Since the uniform persistence of the dynamics has been proven in Lemma \ref{lemma-persistent}, we only need to prove that the dynamics satisfies Conditions (I) and (II) in Theorem \ref{thm-pre-global-attractivity}. 
	
	From Property \ref{pro-subhomo} and Theorem \ref{thm-pre-subhomogeneous}, we know $f(i,y)$ is strongly subhomogeneous. For any $\eta\in(0,1)$, let $u_1(t) = \eta \psi(t,i_0,y)$ and $u_2(t) = \psi(t,\eta i_0,y)$. Then, $u_1(t)$ satisfies 
	\begin{align*}
	\frac{du_1(t)}{dt} = \eta \frac{d{\psi}(t,i_0,y)}{dt} = \eta f(\psi(t,i_0,y),y)\\
	\le f(\eta\psi(t,i_0,y),y) 
	= f(u_1(t),y) 
	\end{align*}
	and $u_2(t)$ satisfies $du_2(t)/dt = f(u_2(t),y)$. From the comparison theory of differential equations and $u_1(0) = u_2(0)=\eta i_0$, we have that $u_1(t)\le u_2(t)$, i.e. $\psi(t,\eta i_0,y)\ge \eta \psi(t,i_0,y)$ for $\forall t,i_0,y$. Hence, we know that for any $t\in\R$ and $y\in Y$, $\psi(t,\cdot,y)$ is subhomogeneous on $[0,1]^n$. Recall that we have proved that $\psi(t,i_0,y)$ is uniformly persistent, namely that there exists $T_2$ such that $\psi(t,i_0,y)\gg \mathbf{0}$ for $t\ge T_2$ and any $i_0,y$. By the strong subhomogeneity of $f(\cdot,y)$, it follows that $u_1(t_0)\ll u_2(t_0)$ for some $t_0\ge T_2$, implying that for any $y\in Y$, $\psi(t_0,\cdot,y)$ is strongly subhomogeneous on $[0,1]^n$. 
	
	Note that Theorem 4.1.1 in \cite{smith2008monotone} proved the monotonicity for cooperative and irreducible dynamical systems.  Here we extend this result to our model (\ref{eq:unified-generic-model}). For each $(i,y)\in [0,1]^n\times Y$, let $Z(t) = \partial \psi(t,i,y)/\partial i $. Then, we have 
	\begin{align}\label{linear-sys}
	\frac{dZ(t)}{dt} = B(t) Z(t), ~{\rm where}~B(t) = D_i f(i,y).
	\end{align}
	It follows from Lemma \ref{lemma-linear-sys} that every element of $Z(t)$ is nonnegative. Then, the monotonicity of $\psi(t,\cdot,y)$ follows from 
	\begin{align} \nonumber 
	& \psi(t,\hat{i}_0,y)-\psi(t,i_0,y) \\ \label{mono-express}
	= &\int_{0}^1\frac{\partial \psi}{\partial i}(t,i_0+r(\hat{i}_0-i_0),y) (\hat{i}_0-i_0) dr.
	\end{align}
	
	For any $i_0,\hat{i}_0\gg \mathbf{0}$, we can always find $\eta\in (0,1)$ such that 
	$\eta i_0 \le \hat{i}_0 \le \eta^{-1} i_0$.
	Then, by the subhomogeneity and monotonicity of  $\psi(t,\cdot,y)$ on $[0,1]^n$, we know that for any $t\in\R$,
	\begin{align*}
	\eta \psi(t,i_0,y) \le \psi(t,\hat{i}_0,y) \le \eta^{-1} \psi(t,i_0,y).
	\end{align*}
	From the strongly subhomogeneity of $\psi(t_0,\cdot,y)$, it follows 
	\begin{align*}
	\eta \psi(t_0,i_0,y) \ll \psi(t_0,\hat{i}_0,y) \ll \eta^{-1} \psi(t_0,i_0,y).
	\end{align*}
	This means Conditions (I) and (II) in Theorem \ref{thm-pre-global-attractivity} are satisfied.
	This completes the proof of part (i).
	
	Now we prove part (ii). From $\alpha_v(t)=0$, $h_{v}(i(t),\beta_v(t))\le 1$, $g_v(i(t),\alpha_v(t),\Gamma(t))\ge 0$ for $\forall v\in V$ and $t\ge 0$, we know
	$\frac{di_{v}(t)}{dt} \ge - i_v(t)$ always holds, implying $i_v(t)\ge \exp(-t)i_v(0)$. It follows that when $i_v(0)>0$, $i_v(t)>0$ holds for any $t\ge 0$. From the strong connectivity of $\mathcal G(M(\Gamma))$ and Property \ref{pro-mono}, we know there exists $T_1>0$ such that $i_v(T_1)>0$ for any $v\in V$. Therefore, for any $i_0, \hat{i}_0\in [0,1]^n\bo$, we can find $\eta\in(0,1)$ such that 
	\begin{align}\label{mono-1}
	\eta \psi(T_1,i_0,y) \le \psi(T_1,\hat{i}_0,y)\le  \eta^{-1} \psi(T_1,i_0,y).
	\end{align}
	By combing with the subhomogeneous and monotonicity of  $\psi(t,\cdot,y)$ on $[0,1]^n$, we have that for $t\ge T_1$,
	\begin{align}\label{mono-2}
	\eta\psi(t,i_0,y) \le \psi(t,\hat{i}_0,y) \le \eta^{-1} \psi(t,i_0,y).
	\end{align}
	From the strong  subhomogeneity of solution $\psi(t_0,\cdot,y)$ proven in the preceding part (i), we have that for any $n\ge 1, n\in \N$,
	\begin{align}\nonumber
	\eta\psi(nT_0+T_1,i_0,y)\ll &\psi(nT_0+T_1,\hat{i}_0,y) \\ \label{mono-3}
	\ll &\eta^{-1} \psi(nT_0+T_1,i_0,y).
	\end{align}
	We recall the metric $\rho(\cdot,\cdot)$ given in \cite{ZhaoBook2003}: 
	\begin{align*}
	\rho(i_0,\hat{i}_0) = \inf\{\ln\eta~|\eta\ge 1, ~\eta^{-1} i_0\le \hat{i}_0\le \eta i_0\},\\~~\forall i_0,\hat{i}_0\in (0,1]^n.
	\end{align*}
	
	Let $X = [0,1]^n$. For any $y\in Y$, we define $\hat{\rho}: X\to \mathbb R^+$ as $$\hat{\rho}(i_0,\hat{i}_0) = \rho (\psi(T_1,i_0,y),\psi(T_1,\hat{i}_0,y))$$ 
	for $\forall i_0,\hat{i}_0\in X\bo$, $\hat{\rho}(i_0,\mathbf{0})=+\infty$ and $\hat{\rho}(\mathbf{0},\mathbf{0})=0$. It can be seen that $(X,\hat{\rho})$ is a metric space and $\hat{\rho}(\cdot,\cdot)$ is continuous with respect to the product topology induced by the norm $\|\cdot\|$. 
	Moreover, inequalities (\ref{mono-1}) and (\ref{mono-2}) imply that for any $i_0,\hat{i}_0\in X\bo$ and $t\ge T_1$,
	\begin{align}\label{mono-new-metric}
	\hat{\rho}(\psi(t,i_0,y),\psi(t,\hat{i}_0,y)) \le \hat{\rho}(i_0,\hat{i}_0), 
	\end{align}
	while inequalities (\ref{mono-1}) and (\ref{mono-3}) imply that for any $i_0, \hat{i}_0 \in X\bo$ with $i_0\neq \hat{i}_0$ and $n\ge 1, n\in \N$,
	$$
	\hat{\rho}(\psi(n T_0,+T_1,i_0,y),\psi(nT_0,+T_1,,\hat{i}_0,y)) < \hat{\rho}(i_0,\hat{i}_0).
	$$
	
	Since $X$ and $Y$ are compact, we have that any trajectory  $(\psi(t,i_0,y),\theta(t,y))$ has its $\omega$-limit set $K_0\subseteq X\times Y$. 
	
	If $K_0|_{X} \neq \{\mathbf{0}\}$, similar to the proof of Theorem 2.3.5 in \cite{ZhaoBook2003} (Theorem \ref{thm-pre-global-attractivity} in the present paper), we can prove that for any $y_0\in K_0|_{Y}$, the cardinality of set $p^{-1}(y_0)\cap K_0$ is one, i.e., there is only one point $i^* \in K_0|_{X}$ satisfies $(i^*,y_0)\in K_0$. For any trajectory with any nonzero initial value, its $\omega$-limit set equals $K_0$. This implies that for any $y\in Y$, model \eqref{eq:unified-generic-model} is globally attractive.
	
	If there exists $y_0\in Y$ such that $\psi(t,i^*_{y_0},y_0)\in (0,1]^n$, then $K_0|_{X} \subseteq (0,1]^n$. Notice that for any trajectory with any nonzero initial value, its $\omega$-limit set equals $K_0$, implying that $\psi(t,i^*_{y},y)\in (0,1]^n$ holds for almost every $y\in Y$.
	
	If $K_0|_{X} = \{\mathbf{0}\}$, then by inequality (\ref{mono-new-metric}) and the assumption that $\hat{\rho}(\mathbf{0},i_0)=+\infty$, we have that for any trajectory with any nonzero initial value, its omega limit set equals $K_0$. This implies that the equilibrium $\mathbf{0}$ is globally attractive.
	
	Finally, part (iii) follows from Theorem \ref{thm-alpha-0-stability} immediately. 
\end{proof}

\section{Proof of Lemma \ref{lemma-alpha>0}}
\label{appendix:lemma-alpha>0}

\begin{proof}
	From $h_{v}(i(t),\beta_v(t))\le 1$ and $g_v(i(t),\alpha_v(t),\Gamma(t))\ge \alpha_v(t)$, we know that for $t\ge 0$,
	\begin{align*}
	\frac{di_{v}(t)}{dt} \ge - i_v(t) + \alpha_{v}(t)\left(1-i_v(t)\right)\ge -2 i_v(t) + \alpha_v(t).
	\end{align*}
	It follows from $M(\alpha_v)>0$ that there exist $\delta_1>0$ and $T_0>0$ such that
	$$
	\int_{kT_0}^{(k+1)T_0}\alpha_{v_0}(s)ds\ge \delta_1 
	$$
	holds for any $k\in\N$. Then for $t\in[NT_0, (N+1)T_0)$ and $N\in \mathbb N$, we have
	\begin{align*}
	i_{v_0}(t)\ge  &e^{-2t}i_{v_0}(0) + \int_{0}^t e^{2(s-t)}\alpha_{v_0}(s)ds \\
	\ge  &\sum_{k=0}^Ne^{-2t}\int_{2k T_0}^{(k+1)T_0} e^{k T_0}\alpha_{v_0}(s)ds \\
	\ge & \frac{e^{-2T_0}-e^{-2(N+1)T_0}}{e^{2T_0}-1}\delta_1.
	\end{align*}
	Therefore, the trajectory of node $v_0$ is persistent. Since $\mathcal G(M(\Gamma))$ is strongly connected, we know that there exist paths from $v_0$ to other nodes in $\mathcal G(M(\Gamma))$. Then, we have that there exist node $u_0\in V$, $\delta>0$ and $T>0$ such that $\int_{t_{k}}^{t_{k+1}}\gamma_{u_0v_0}ds \ge \delta$, where $t_k = kT, k\in \N$. It then follows from Property \ref{pro-mono} that there exists $\delta'>0$ such that for any $k\in\N$, 
	\begin{align}\label{partial-u-v-est}
	\int_{t_{k}}^{t_{k+1}}\frac{\partial g_{u_0}}{\partial i_{v_0}}(\cdot,\alpha_{v_0}(s),\Gamma(s)) ds \ge \delta'.
	\end{align}
	Then, from $h_{u_0}\le 1$, $\partial g_v/\partial i_u\ge 0$ and the integral mean value theorem, we have
	\begin{align*}
	\frac{d{i}_{u_0}(t)}{dt} \ge &-h_{u_0}i_{u_0} + g_{u_0} (1-i_{u_0})\\\nonumber
	= & -(h_{u_0}+g_{u_0})i_{u_0} +\bigg[\alpha_{u_0} + \frac{\partial g_{u_0}}{\partial i_{v_0}}(\hat{i},\alpha_{v_0},\Gamma)i_{v_0} \\
	&~~~~~~~~~~~~~~~~~~~~~~~~~+\sum_{v\neq v_0} \frac{\partial g_{u_0}}{\partial i_{v}}(\hat{i},\alpha_{v_0},\Gamma)i_{v} \bigg] \\
	\ge &-2i_{u_0} + \alpha_{u_0} + \frac{\partial g_{u_0}}{\partial i_{v_0}}(\hat{i},\alpha_{v_0},\Gamma)i_{v_0}. 
	\end{align*}
	By combing with Inequality (\ref{partial-u-v-est}), we can see that there exists $T_{u_0}$ such that $i_{u_0}(t)\ge \delta_{u_0}$ for $t\ge T_{u_0}$. By induction, we can prove that for any $v\in\{\oneton\}$, there exists $\delta''>0$ and $T'>0$ such that $i_{v}(t)\ge \delta''$. In other words, all trajectories are uniformly persistent. On the other hand, it follows from the proof of Lemma \ref{lemma-alpha-0-ergodic} that  $\psi(t,\cdot,y)$ satisfies Conditions (I) and (II) in Theorem \ref{thm-pre-global-attractivity}, meaning that model \eqref{eq:unified-generic-model} is
	globally attractive.
\end{proof}

\section{Proof of Lemma \ref{lemma-disconnect}}
\label{appendix:lemma-disconnect}

\begin{proof}
	First, we prove part (i) of Lemma \ref{lemma-disconnect}. The global attractivity for the dynamics corresponding to strongly connected component $SCC_1$ can be derived from Lemmas \ref{lemma-alpha-0-ergodic} and \ref{lemma-alpha>0} directly. Let $\phi(t,i^*_{SCC_1},y)$ denote the globally attractive trajectory corresponding to $SCC_1$. For any given ergodic function $z(t)\in [0,1]^{|SCC_1|}$, the system
	\begin{align}\label{dyn-scc-1}
	\frac{d i_{v}(t)}{dt} = f_v(\left[z(t),[i(t)]_{SCC_2}\right],y),~v\in SCC_2
	\end{align}
	has a globally attractive trajectory in $[0,1]^{|SCC_2|}\bo$ by treating $(y(t),z(t))$ as enlarged parameters.
	
	Let $\{\epsilon_k\}_{k\in\N}$ be a positive and monotone decreasing sequence that converges to 0. Then, by the global attractivity of $SCC_1$, we have that for any $i_{2,0}\in[0,1]^{|SCC_2|}$ and any $\epsilon_k$, there exists $T_k>0$ such that for $t>T_k$ and any $i_{1,0}\in[0,1]^{|SCC_1|}$,
	\begin{align*}
	\max\{\mathbf{0}_{SCC_1},\phi(t,i^*_{SCC_1},y)- \epsilon_k \mathbf{1}_{SCC_1}\} \\
	\le [\psi(t,[i_{1,0},i_{2,0}],y)]_{SCC_1} \le \phi(t,i^*_{SCC_1},y) + \epsilon_k \mathbf{1}_{SCC_1}.
	\end{align*}
	For brevity, denote 
	\begin{eqnarray*}
		i(t) &=& \psi(t,[i_{1,0},i_{2,0}],y),\\ i_{k,1}^l(t) &=& \max\{\mathbf{0}_{SCC_1},\phi(t,i^*_{SCC_1},y)- \epsilon_k \mathbf{1}_{SCC_1}\},\\
		i_{k,1}^u(t) &=& \phi(t,i^*_{SCC_1},y) + \epsilon_k \mathbf{1}_{SCC_1}.
	\end{eqnarray*}
	Let $i_{k,2}^q(t)$ denote the globally attractive trajectory of system (\ref{dyn-scc-1}) with $z(t) = i_{k,1}^q(t)$, $q = l,u$, respectively. According to Theorem 1.7 in \cite{HirschPaper1985}, a cooperative function is monotone, meaning that $f_v(i,y)$ is monotone in $i$, which implies 
	\begin{align*}
	f_v\left(\left[i_{k,1}^l(t),[i(t)]_{SCC_2}\right],y\right)\le f_v\left(i(t),y\right)\\
	\le f_v\left(\left[i_{k,1}^u(t),[i(t)]_{SCC_2}\right],y\right).
	\end{align*}
	Then it can be concluded that $i_{k,2}^l(t)\le [i(t)]_{SCC_2}\le i_{k,2}^u(t)$. From $\lim_{k\to\infty}\|i_{k,1}^u(t)-i_{k,1}^l(t)\| = 0$ and $f$ is continuously differentiable, we can see that  $\lim_{k\to\infty}\|i_{k,2}^u(t)-i_{k,2}^l(t)\|=0$. Therefore, there exists $i^*$ such that
	$$
	\lim_{k\to\infty}i_{k,2}^l(t) =  \lim_{k\to\infty}i_{k,2}^u(t) =  [\psi(t,i^*,y)]_{SCC_2}
	$$
	for any $i_{1,0}\in[0,1]^{|SCC_1|}$ and $[\psi(t,i^*,y)]_{SCC_2} = \phi(t,i^*_{SCC_1},y)$. That is, $\psi(t,i^*,y)$ is the globally attractive trajectory we are seeking for $SCC_1\cup SCC_2$.  
	
	Let $V_{\alpha = 0} = \{v: \alpha_v(t)=0,~\forall t\in\R\}$ denote the set of nodes that are not subject to pull-based attacks, meaning that $M(\alpha_v)= 0 $. Therefore, $V$ can be partitioned into two sets such that $V = V_{\alpha=0}\cup V_{\alpha>0}$. This means that $\V_{SCC_2}\cap V_{\alpha>0} = \emptyset$ is equivalent to $\V_{SCC_2}\subseteq V_{\alpha=0}$, namely that $\alpha_v(t) = 0 $, $\forall v\in \V_{SCC_2}$.
	When $\phi(t,i_{SCC_1}^*,y) = \mathbf{0}_{|SCC_1|}$, $\V_{SCC_2}\cap V_{\alpha>0} = \emptyset$ and $\mu(Df_{22})< 0$, the global attractivity trajectory corresponding to $SCC_2$, i.e., $[\psi(t,i^*,y)]_{SCC_2}$ satisfies system (\ref{dyn-scc-1}) with $z(t) = \mathbf{0}_{|SCC_1|}$. Then it follows from part (iii) of Lemma \ref{lemma-alpha-0-ergodic} that $SSC_2$ globally converges to $\mathbf{0}_{|SCC_2|}$. This completed the proof of proves part (i) of Lemma \ref{lemma-disconnect}. 
	
	Now we prove part (ii) of Lemma \ref{lemma-disconnect}.
	If $\V_{SCC_2}\cap V_{\alpha>0} \neq \emptyset$ or $\phi(t,i_{SCC_1}^*,y) \in (0,1]^{|SCC_1|}$, it follows from Lemma \ref{lemma-alpha>0} that $[\psi(t,i^*,y)]_{SCC_2} \in (0,1]^{|SCC_2|}$ is globally attractive in $[0,1]^{|SCC_2|}$. 
	If $\phi(t,i_{SCC_1}^*,y) = \mathbf{0}_{|SCC_1|}$, $\V_{SCC_2}\cap V_{\alpha>0} = \emptyset$ and $\mu(Df_{22})\ge 0$, by substituting $z(t) = \mathbf{0}_{SCC_1}$ into system (\ref{dyn-scc-1}) and employing Lemma \ref{lemma-alpha-0-ergodic}, we have that $[\psi(t,i^*,y)]_{SCC_2} \in [0,1]^{|SCC_2|}$ is globally attractive in $[0,1]^{|SCC_2|}\bo$. 
	If $\phi(t,i_{SCC_1}^*,y) \notin \{\mathbf{0}_{|SCC_1|}\}\cup (0,1]^{|SCC_1|}$ and $\V_{SCC_2}\cap V_{\alpha>0} = \emptyset$, similar to the proof of Lemmas \ref{lemma-alpha-0-ergodic} and \ref{lemma-alpha>0}, we can prove that there exists a globally attractive trajectory $[\psi(t,i^*,y)]_{SCC_2} \in [0,1]^{|SCC_2|}$ for $SCC_2$.
	This completes the proof of part (ii) of Lemma \ref{lemma-disconnect}. 
\end{proof}

\section{Proof of Theorem \ref{thm-general}}
\label{appendix:thm-general}
\begin{proof}
	We can always divide the graph $\mathcal G(M(\Gamma))$ into $K$ strongly connected components, denoted by $SCC_1,\cdots,SCC_K$. Let $V_{SCC_{k}}$ denote the set of nodes in $SCC_k$ and $|SCC_k|$ the number of the nodes in $SCC_k$, $k=1,\cdots,K$. Let $R_k$ be the indices of $SCCs$ that have links pointing to $SCC_k$. This means that if $R_k\neq \emptyset$ and $j\in R_k$, there exists a node in $SCC_j$ that has a path to a node in $SCC_k$. From Properties \ref{pro-mono} and $\partial f_v(\mathbf{0},y)/{\partial i_v} =\partial g_v(\mathbf{0},\alpha_v,\Gamma)/{\partial i_v}$, we know that the Jacobian matrix $D_i f(\mathbf{0},y)$ has the Perron-Frobenius form \eqref{Df-PF-form-K}. Following the argument in the proof of Lemma \ref{lemma-disconnect}, we can use induction to prove that for each $SCC_k$, 
	\begin{itemize}
		\item $\mathbf{0}_{|SCC_k|}$ is globally attractive in $[0,1]^n$ if either of the following condition holds:
		\begin{itemize}
			\item $\V_{SCC_k}\cap V_{\alpha>0}= \emptyset$, $\mu(Df_{kk})< 0$ and ${R_k}= \emptyset$;
			\item $\V_{SCC_k}\cap V_{\alpha>0}= \emptyset$, $\mu(Df_{kk})< 0$, ${R_k}\neq \emptyset$ and $\mathbf{0}_{|SCC_r|}$ is globally attractive for every $r\in {R_k}$.
		\end{itemize}
		\item $SCC_k$ has a positive trajectory that is globally attractive in $[0,1]^{|SCC_k|}\bo$ if one of the following condition holds:
		\begin{itemize}
			\item  $\V_{SCC_k}\cap V_{\alpha>0}= \emptyset$,  $\mu(Df_{kk})> 0$ and $R_k =\emptyset$;
			\item  ${R_k}\neq \emptyset$ and there exists $j\in R_k$ such that $SCC_j$ has a positive and globally attractive trajectory.
		\end{itemize}
		\item  $SCC_k$ has a positive trajectory that is globally attractive in $[0,1]^{|SCC_k|}$ if $\V_{SCC_k}\cap V_{\alpha>0}\neq  \emptyset$.
		\item  $SCC_k$ has a trajectory that is globally attractive in $[0,1]^{|SCC_k|}\bo$ if $\V_{SCC_k}\cap V_{\alpha>0}=  \emptyset$ and $\mu(Df_{kk})\ge 0$.
	\end{itemize}
	Therefore, it can be concluded that for $V = \cup_{k=1}^K \V_{SCC_k}$, there exists a globally attractive trajectory $\psi(t,i^*,y)\in [0,1]^{n}\bo$. Moreover, following the argument in the proof of Lemma \ref{lemma-alpha>0}, we know that if node $v$ is subject to pull-based attacks, the trajectory of node $v$ is uniformly persistent, implying $[\psi(t,i^*,y)]_v\neq 0$.  
\end{proof}

\section{Proof of Theorem \ref{thm-almost-periodic-global}}
\label{appendix:thm-almost-periodic-global}
\begin{proof}
	For the almost periodic function $y = \{y(t)\}_{t\in\R}$, let $Y_0$ be the closure of $\{\theta(s,y), s\in\R\}$ and $\mathcal F_0$ be the Borel $\sigma$-algebra of $Y_0$. The normalized Haar measure of $Y_0$, denoted by $\mathbb P_0$, is the unique $\theta$-invariant and ergodic probability measure \cite{arnold2013random}. By Definition \ref{def-ergodic}, $\{y(t)\}_{t\in\mathbb R}$ is ergodic with respect to the probability space $(Y_0,\F_0,\mathbb P_0)$. Then it follows from Theorem \ref{thm-general} that for almost every $y_0\in Y_0$, there is a globally attractive trajectory in $[0,1]^n\bo$. Suppose for $y_1 = \theta(t_1,y)\in Y_0$, there is a globally attractive trajectory $\psi(t,i_1^*,y_1)$, namely that $\lim_{t\to\infty}\|\psi(t,i_0,y_1)-\psi(t,i^*_1,y_1)\|=0$ for $\forall i_0\in[0,1]^n\bo$. Let $i^* = \psi(-t_1,i^*_1,y_1)$. Then,
	$$
	\psi(t_1,i^*,y) = \psi(t_1,\psi(-t_1,i^*_1,y_1),y) = \psi(0,i^*_1,y_1) = i^*_1.
	$$ 
	For any $i_0\in[0,1]^n\bo$ and $t> t_1$, we have
	\begin{align*}
	&\|\psi(t,i_0,y)-\psi(t,i^*,y)\|\\ 
	=&\|\psi(t-t_1,\psi(t_1,i_0,y),y_1)-
	\psi(t-t_1,i_1^*,y_1)\|,
	\end{align*}
	implying that $\psi(t,i^*,y)$ is globally attractive. 
	
	Now we prove the existence of an almost periodic solution. From almost periodicity of $y(t)$ (see Definition \ref{def-almost-periodic}), it follows that there exists $l>0$ such that any $l$-length interval contains a constant $\xi>0$ so that $\|y(t+\xi)-y(t)\|\le \epsilon$ holds for all $t\in \R$, namely $\xi$ is an $\epsilon$-translation number of $y(t)$. By the global attractivity of model (\ref{eq:unified-generic-model}) and the almost periodicity of $y(\cdot)$, it follows that for any constant $M>0$, there exists $T>0$ such that for any $t>T$,
	\begin{align*}
	&\|\psi(t+\xi,i,y)-\psi(t,i,y)\|\\
	=&\|\psi(t,\psi(\xi,i,y),\theta(\xi,y))-\psi(t,i,y)\|\\ 
	\le&\|\psi(t,\psi(\xi,i,y),\theta(\xi,y))-\psi(t,i,\theta(\xi,y))\|\\
	&+\|\psi(t,i,\theta(\xi,y))-\psi(t,i,y)\|\\
	\le &M\epsilon.
	\end{align*}
	That is, $\psi(t,i,y)$ is asymptotically almost periodic. 
	
	Taking $\{t_{k}\}_{k\in\mathbb N}$ such that $$\sup_{t,i}|f(i,y(t))-f(i,y(t+t_{k}))|<1/k$$ and considering the solution $\psi(t+t_k,i,y)$ to $d{i}(t)/dt=f(i,y(t+t_{k})), \forall k\in \N$. According to Lemma 2 in \cite{lu2008almost}, we know that there exists a sub-sequence of $\{\psi(t+t_{k},i,y)\}_{k\in\N}$, denoted by $\psi(t+t_{k},i,y)$ for not to introduce extra notations, that converges uniformly and its limit at $k\to\infty$, denoted by $\psi^{*}(t,i,y)$, is a solution to model (\ref{eq:unified-generic-model}). Since  $\psi(t+t_{k},i,y)$ is asymptotically almost periodic, i.e., $|\psi(t+t_{k}+\xi,i,y)-\psi(t+t_{k},i,y)|\le M\epsilon$ for sufficiently large $k$, it follows that the limit $\psi^{*}(t,i,y)$ is almost periodic in $t$ and $\xi$ is an $M\epsilon$-translation number of $\psi^{*}(t,i,y)$.
\end{proof}

\section{Proof of Theorem \ref{thm-bounds}}
\label{appendix:thm-bounds}

\begin{proof}
	For each $v\in V$, $i_v(t)$ in model \eqref{eq:unified-generic-model} satisfies
	\begin{align*}
	\frac{d i_v(t)}{dt} 
	=& -h_v\left(i(t),\beta_v(t)\right)\cdot i_v(t)\\ 
	&+g_v(i(t),{\alpha}_v(t),{\Gamma}(t))\cdot\left(1-i_v(t)\right)\\
	\ge &-\overline{h}_v(\overline{\beta}_v)i_v(t) + g_v(i_{\min},\underline{\alpha}_v,\underline{\Gamma})(1-i_v(t))\\
	= &-[\overline{h}_v(\overline{\beta}_v) + g_v(i_{\min},\underline{\alpha}_v,\underline{\Gamma})] i_v(t) + g_v(i_{\min},\underline{\alpha}_v,\underline{\Gamma})
	\end{align*}
	and 
	\begin{align*}
	\frac{d i_v(t)}{dt} 
	\le & -\underline{h}_v(\underline{\beta}_v)i_v(t) + g_v(i_{\max},\overline{\alpha}_v,\overline{\Gamma})(1-i_v(t))\\
	= &-[\underline{h}_v(\underline{\beta}_v)+g_v(i_{\max},\overline{\alpha}_v,\overline{\Gamma})] i_v(t) + g_v(i_{\max},\overline{\alpha}_v,\overline{\Gamma}).
	\end{align*}
	By the Gronwall inequality \cite{Robinson}, we obtain the bounds of $i_v(t)$. This completes the proof.
\end{proof}

\end{document}